\newif\ifsubmode
\submodefalse
\submodetrue

\documentclass{emulateapj}
\usepackage{lscape}

\ifsubmode
\else
\slugcomment{\today}
\fi

\def\loneosI{LONEOS~-~I\ }
\def\loneosII{LONEOS~-~II\ }
\def\gsim{\,\lower3pt\hbox{$\sim$}\llap{\raise2pt\hbox{$>$}}\,}
\newcommand{\Chisqr}{$\chi^2\ $}
\newcommand{\Chisqrm}{\chi^2}
\newcommand{\etal}{{et al.~}}

\shorttitle{LONEOS-I RR Lyrae Stars}
\shortauthors{Miceli \etal}

\begin{document}

\title{Evidence for Distinct Components of the Galactic Stellar 
Halo from 838 RR Lyrae Stars Discovered in the LONEOS-I Survey}

\author{Antonino Miceli}
\affil{Astronomy and Physics Departments, University of Washington, Box 351580, Seattle, WA 98195-1580 USA,\\ {\it amiceli@astro.washington.edu} }

\author{Armin Rest}
\affil{Cerro Tololo Inter-American Observatory, National Optical Astronomy Observatory, Casilla 603, LaSerena, Chile, \\ {\it arest@ctio.noao.edu} }

\author{Christopher W. Stubbs}
\affil{Department of Physics and Harvard-Smithsonian Center for Astrophysics, Harvard University, 17 Oxford Street, Cambridge, MA 02138 USA }

\author{Suzanne L. Hawley}
\affil{Astronomy Department, University of Washington, Box 351580,  Seattle, WA 98195-1580 USA}

\author{Kem H. Cook}
\affil{Lawrence Livermore National Laboratory, 7000 East Ave., Livermore, CA 94550 USA}

\author{Eugene A. Magnier}
\affil{University of Hawaii, Institute for Astronomy, 2680 Woodlawn Dr., Honolulu, HI 96822 USA}

\author{Kevin Krisciunas}
\affil{University of Notre Dame, Department of Physics, 225 Nieuwland Science Hall, Notre Dame, IN  46556 USA}

\author{Edward Bowell}
\affil{Lowell Observatory, 1400 West Mars Hill Road, Flagstaff, AZ 86001 USA}

\and

\author{Bruce Koehn}
\affil{Lowell Observatory, 1400 West Mars Hill Road, Flagstaff, AZ 86001 USA}

 
\begin{abstract}
We present 838 {\it ab}-type RR Lyrae stars from the Lowell Observatory Near 
Earth Objects Survey Phase I (LONEOS-I). These objects cover 1430 $deg^{2}$ 
and span distances ranging from 3-30kpc
from the Galactic Center. Object selection is based on phased, photometric data with
28-50 epochs. We use this large sample to explore the bulk properties of the stellar halo,
including the spatial distribution. The period-amplitude distribution of this sample shows
that the majority of these RR Lyrae stars resemble Oosterhoff type I, but there is a significant fraction
(26\%) which have longer periods and appear to be Oosterhoff type II. We find that
the radial distributions of these two populations have significantly different
profiles ($\rho_{OoI} \sim R^{-2.26 \pm 0.07}$
and $\rho_{OoII} \sim R^{-2.88 \pm 0.11}$). This suggests that the stellar
halo was formed by at least two distinct accretion processes and supports 
dual-halo models. 
\end{abstract}



\section{Introduction}
A major goal of modern astrophysics is to understand the physical
processes by which galaxies form, and to place galaxy formation in the
context of models for the evolution of large-scale structure in the
Universe. The advent of wide-field CCD cameras has now made it
possible to survey large samples of Galactic stellar populations, and
therefore to study the structure of our Galaxy as a whole.  While
keeping in mind that the Galaxy may not always be representative of
(spiral) galaxies in general, the study of our Galaxy is now a
fruitful avenue for the detailed exploration of galactic structure,
providing the empirical basis for differentiating between formation
models.

RR Lyrae stars are horizontal branch stars which lie in the
instability strip.  They represent a late stage of stellar evolution,
and as such form an old stellar population that can be used to study
the early stages of galaxy formation.
RR Lyrae stars  have a long history as tracers of Galactic structure
\citep{Kinman66,Oort75,Hawkins84,Saha85}. They are also relatively
easy to identify from their colors and variability, enabling the
formation of samples with little contamination.  \citet{Suntzeff91}
estimate the number of field RR Lyrae in the Galactic halo to be about
85,000, but less than 8,000 have been discovered, including those in
the bulge and thick disk \citep[e.g.,][]{Heck88}. Further, their
absolute magnitudes (for RR Lyrae {\it ab}-types) have been measured
to better than 0.1 magnitudes using robust statistical parallax
\citep{Layden96} and direct parallax techniques \citep{Benedict02}.
However, RR Lyrae stars do have a disadvantage. The relative abundance
of RR Lyrae stars depends on the horizontal branch morphology of the
RR Lyrae progenitor system. Thus, they may not reveal the entire
history of Galactic formation. Overall, RR Lyrae stars are an incisive
probe of the halo field population and may provide useful insights on
Galactic formation. In recent years, several sky survey have discovered large numbers of
Galactic RR Lyrae over large fractions of the sky, e.g, SDSS
\citep{Ivezic00}, QUEST \citep{Vivas2006}, 
and NSVS \citep{Kinemuchi2006}. The SDSS and QUEST surveys have shown that
RR Lyrae can be used to find and trace the remants of accreted satellite 
galaxies in the Galaxy.  

RR Lyrae also have been well cataloged and studied in the globular 
clusters \citep[e.g.,][]{Clement2001}. 
In particular, globular clusters fall into two broad groups based on the mean periods
of their RRab stars and on their RRc fraction, with Oosterhoff I (OoI)  clusters having shorter
mean periods and lower RRc content compared to Oosterhoff II (OoII) clusters
\citep{Oosterhoff}. 
In addition, these two classes populate distinct regions in period-amplitude space.  
There are certain stellar systems (most Local Group dwarf spheroidals) which cannot be
classified as OoI nor OoII, but rather exhibit period-amplitude
distributions which fall between OoI and OoII. These are referred to
as intermediate.  We can examine Galactic stellar halo RR Lyrae 
in a similar manner to determine their Oosterhoff classification.
\citet{Suntzeff91} found OoI and OoII components in field RR Lyrae stars, 
but no intermediate Oo component (i.e., the Oosterhoff ``gap'').
The origin of this Oosterhoff effect has been studied intensively
\citep[e.g.,][]{Sandage1981, Lee1990, Lee99}. 
The Oosterhoff classification may either indicate an age 
difference or a metallicity difference. 
Regardless, the Oosterhoff classification of the RR
Lyrae in the Galactic stellar halo will provide insights into it formation
history.

In this paper, we present new results from the
\loneosI RR Lyrae survey.  Using 838 RR Lyrae stars discovered from
the \loneosI survey, we will investigate their distribution in the
stellar halo out to distances of 30 kpc, as a function of Galactic
position.  In Section 2, we introduce the \loneosI variability survey
and discuss the photometric pipeline. In Section 3, we present the
\loneosI RR Lyrae sample and discuss the selection process and
detection efficiency. In Section 4, we discuss the properties of the
\loneosI RR Lyrae stars. Finally, we discuss the implications for
Galactic formation and look to the future.

\section{\loneosI Variability Survey}

The Lowell Observatory Near Earth Object Survey (LONEOS)
\citep{Bowell1995} has as its primary goal the detection of
potentially hazardous asteroids.  Situated on the Anderson Mesa
outside Flagstaff, Arizona, LONEOS is carried out with a 0.6m Schmidt
telescope.  During the first two years of operation, imaging was done
with a camera constructed at the University of Washington
\citep{Diercks1995}, with a single 2k x 4k unfiltered CCD (5
sq. deg. FOV with $2.67''$/pixels) (\loneosI).  
Under optimal conditions, LONEOS can image up to 1000 square degrees per night,
multiple times. The \loneosI survey imaged fields three times a
night with 60 second exposrure times. 
Fields are re-observed at
least one other time about one lunation later.
\loneosI camera
uses no bandpass filter, but the data can be transformed to
a standard photometric system using external datasets. The \loneosI camera
can reach a depth of $R \sim 18.5$.

Besides searching for moving objects, this dataset can be used to
search for temporal variability, specifically of stellar objects.
Data from LONEOS have been used to construct a stellar temporal
variability database, from which we extracted candidate RR Lyrae
stars.  The cadence of LONEOS observations makes them well suited to
detect short-term variability (e.g., RR Lyrae, binaries, CVs), and
long-term variability (e.g., QSOs).

\begin{figure}[t]
\centering
\includegraphics[scale=0.35,angle=90]{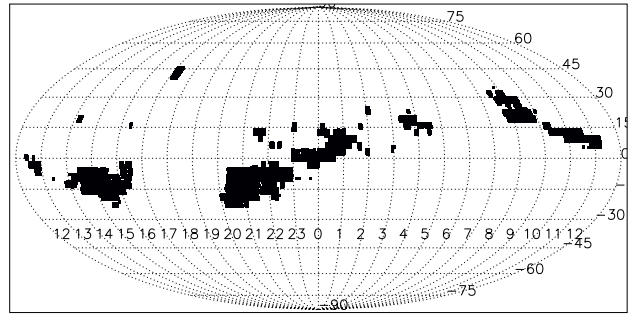}
\caption[\loneosI Sky Coverage]{
The black regions indicate subfields of \loneosI survey data 
with at least 28 good quality images. These fields are
used for our RR Lyrae star analysis in this paper. This area covers
1430 $deg^{2}$ on the sky. The equatorial coordinates of the subfields are
displayed using a Mollweide projection. 
}
\label{fig:skymapOLD}
\end{figure}

The \loneosI photometric variability database discussed in this paper
consists of photometric data obtained between 1998 - 2000. It covers a
large fraction of the sky, much of it imaged in multiple epochs (see
Figure \ref{fig:skymapOLD}). The \loneosI dataset that we will analyze
in this paper consists of 1430 $deg^2$ with at least 28 epochs. 

\subsection{Image Reduction Pipeline}
The \loneosI image reduction pipeline consists of several steps.
Basic CCD processing (bias, flat-field correction) is followed by
object detection using the photometry program DoPhot
\citep{Schechter93}.  Bookkeeping is accomplished by automatic merging
of the DoPhot output file with the FITS header of each parent image to
produce a self-describing and self-contained output file. These output
files are entered into a database (broken down by region on the sky)
where the primary indexing is based on image position (RA, DEC).  The
relative and absolute photometry is performed at the database level.
We divide each field into subfields, and relative photometry is
performed on these subfields.  We tie the \loneosI absolute photometry
to an external catalog. Details are presented in later sections.
Access tools are used to search the database and apply variability
detection algorithms. We have developed a set of robust statistics
which are used to discriminate between long and short term variables
in the timeseries data. It should be noted that we do not attempt to
perform accurate photometry of extended objects. The database should
be strictly used for stellar objects.  For details of the data
reduction pipeline see \citet{Armin_thesis}.

\subsubsection{Photometry}
The photometric calibration can be separated into two independent
parts: relative photometry and absolute photometry.  For
identification and classification of variable objects, relative
photometry suffices. Since LONEOS uses an unfiltered system, we must
rely on external datasets to tie the relative magnitudes to a standard
photometric system. We outline the procedure in this section.

For an image taken at a given time $t$ and airmass $a(t)$ through a filter $X$,
the apparent magnitude $m(t,s)$ of a star $s$ with color $X-Y$
(e.g. $X-Y=B-V$) can be written as
\begin{eqnarray}
  m(t,s)     & = &  m_{instr}(t,s) + A_X(t) + c_{1} * a(t) \nonumber \\ 
  && + c_{2} * (X-Y) * a(t) +  c_{3} * (X-Y) + m_{0}.
\end{eqnarray}
Here $m_{instr}(t,s)$ is the instrumental magnitude with an error of
$\sigma_{instr}(t,s)$, $A_{X}(t)$ is the extinction due to clouds
(assuming 'grey' extinction), $c_{1}$ and $c_{2}$ are the grey and
color-dependent airmass extinction coefficients, respectively. For our
purposes, since we are interested in the photometry of only one type
of star, we can set $(X-Y)$ to a constant, and include $c_{2}$ in $c_{1}$
and $c_{3}$ in $m_{0}$.  LONEOS takes images of fields with generally low
airmass ($76\%$ of images have airmass less than 1.5 with a mean
airmass of 1.3).  Therefore, the second order term depending on color
and airmass is very small anyway. We group all image-dependent terms
into $m_{cal}(t)= A_{X}(t) + c_{1} * a(t)$ and
\begin{eqnarray}
 m(t,s)       & = &  m_{instr}(t,s) + m_{cal}(t) + m_{0} \label{equ:m}.
\end{eqnarray}
A wide field-of-view or strong PSF distortions of the optical system
can cause systematic variations in the photometry across the image.
We divide each LONEOS field (i.e, 2k x 4k) into 50 square {\em
subfields}, each 0.304 degrees on a side. This subfield size is large
enough to include a sufficiently large number of stars to perform the
photometry, but small enough to avoid systematic problems (e.g., Flat
fielding errors, clouds, etc) across the wide FOV of the camera.

\subsubsection{Relative Photometry}

We define the relative magnitude $m_{rel}(t,s)$ and the time-averaged relative
magnitude $\bar m_{rel}(s)$ of the star $s$ as:
\begin{eqnarray}
 m_{rel}(t,s)     & = &  m_{instr}(t,s) + m_{cal}(t), \label{equ:mrel}\\
\bar m_{rel}(s)   & = &  \sum_t \frac{m_{rel}(t,s)}{\sigma_{instr}^{2}(t,s)}, \label{equ:mrelaverage}\\
\bar \sigma_{rel}^{-2}(s) & = & \sum_t \sigma_{instr}^{-2}(t,s).
\end{eqnarray}
We take the following approach to perform relative photometric
calibration: calculate the time-averaged magnitude for each star, then
adjust $m_{cal}(t)$ in each subfield, so that on the
average the deviation from single star measurements to the time-averaged star
magnitude is minimized. The function that is minimized is:
\begin{eqnarray}
\Chisqrm  & = & \sum_{t,s}  (\bar m_{rel}(s)  -  m_{rel}(t,s))^{2}/\sigma^{2}(t,s) +  \nonumber \\ 
&& \sum_t m_{cal}(t),   \\
\sigma^{2}(t,s) & = & \bar \sigma_{rel}^{2}(s) + \sigma_{instr}^{2}(t,s),  \label{equ:sigma}
\end{eqnarray}
where $m_{cal}(t)$ are the free parameters. Note that without the second
term, the $\Chisqrm$ is degenerate to the addition of a constant to the
$m_{cal}(t)$ vector. Therefore we add the second term which is
quadratic in the degenerate direction. Hence our solution has
$\sum_t m_{cal}(t)=0$.  We also add $0.02$ magnitudes in quadrature to
$\sigma_{instr}(t,s)$  to take into account systematic error. This is
intended to compensate for the fact that DoPhot is known to
underestimate photometric errors at the bright end, where the error is
dominated by systematics rather than Poisson noise.  These systematics
include flat fielding errors, PSF variations, and the limited accuracy
of the PSF model. Stars for which the reduced $\Chisqrm_s$ of their 
lightcurve exceeds $2.5$ are flagged as potentially variable and excluded
from the set of stars used to fit $m_{cal}(t)$, with
\begin{eqnarray}
\Chisqrm_s  & = & 1/(N_t-1) \sum_{t}  (\bar m_{rel}(s)  -  m_{rel}(t,s))^{2}/\sigma^{2}(t,s).
\end{eqnarray}
Here $N_t$ is the number of images.  Similarly, we exclude images for
which their reduced $\Chisqrm_t$ exceeds $1.5$, with
\begin{eqnarray}
\Chisqrm_t  & = & 1/(N_s-1) \sum_{s}  (\bar m_{rel}(s)  -  m_{rel}(t,s))^{2}/\sigma^{2}(t,s).
\end{eqnarray}
In addition, we exclude images for which more than $20\%$ of the star
measurements $m_{rel}(t,s)$ deviate by more than 3 $\sigma$ from their
time-averaged magnitude $\bar m_{rel}(s)$. This process is repeated
iteratively until the $m_{cal}(t)$ vector values for each subfield
do not change significantly.

\subsubsection{Absolute Photometry: Tying LONEOS Photometry to
External Datasets}

Since LONEOS uses an unfiltered camera, we must use external datasets to
tie LONEOS photometry to a standard photometric system. 
There exist datasets of calibrated photometry over
large portions of the sky.
We use the Guide Star Catalog
$2.2$ \citep{GSC2} (hitherto, referred to as GSC).  
GSC is complete to 18.5 photographic F-band, and Tycho 2
photometry is used for bright sources.  Thus, the GSC magnitude range
matches well to \loneosI.  The procedure to tie LONEOS relative
photometry to GSC is similar to the relative photometry algorithm. We
restrict the range of color ($-1.1 < (M_{GSC\_J} - M_{GSC\_F}) < 1.5$) for the
GSC stars to be similar to colors of RR Lyrae, which are the focus of
this work.  The spectral response of the \loneosI camera system
matches closest to the GSC F passband. Initially, \loneosI photometry
is tied to GSC F, which we refer to as $M_{LONEOS\_F}$ with error
$\sigma_{LONEOS\_F}$. Combining Equations~\ref{equ:m}, \ref{equ:mrel}
and \ref{equ:mrelaverage}, we define the time-averaged apparent magnitude
$\bar M_{LONEOS\_F}(s)$ of the star $s$ in each subfield $t$ as:
\begin{eqnarray}
\bar M_{LONEOS\_F}(s) & = & \bar m_{rel}(s) + m_{0,GSC\_F}, \\
\bar \sigma(s) & = & \bar \sigma_{rel}^{2}(s) + \sigma_{m_{0,GSC\_F}}^2.
\end{eqnarray}
We determine the free parameter $m_{0,GSC\_F}$ and its error $\sigma_{m_{0,GSC\_F}}$ for a given subfield by minimizing
\begin{eqnarray}
\Chisqrm  & = & \sum_{s}  (M_{GSC\_F}(s) - \bar M_{LONEOS\_F}(s))^{2}/(\sigma_{GSC\_F}^2 + \bar \sigma_{rel}^{2}(s)), 
\end{eqnarray}
where the sum is taken over the stars in a given image.
Once $m_{0}$ is deteremined for each subfield, the LONEOS magnitude tied to GSC F for a given 
star $s$, in that subfield is:
\begin{eqnarray}
M_{LONEOS\_F}(t,s) & = &  m_{instr}(t,s) + m_{cal}(t) +  m_{0,GSC\_F}.
\end{eqnarray}

\begin{figure}[t]
\epsscale{1.0}
\centering
\ifsubmode
\plotone{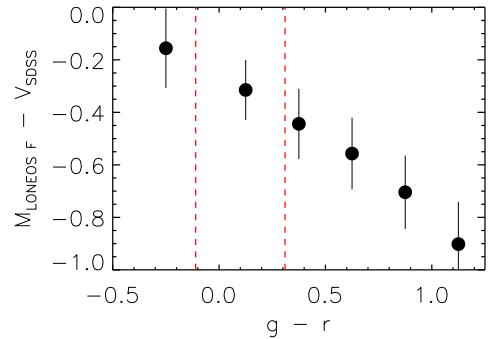}
\fi
\caption{The residual between \loneosI magnitudes (tied to GSC2 F) and 
$V_{SDSS}$ as a function of SDSS color, $g - r$. The two dashed vertical lines 
indicate the color range typical for RR Lyrae. In this range, we see that there is an offset,                                    
$M_{LONEOS\_F} - V_{SDSS} = -0.32$, with an RMS of $0.13$.}
\label{fig:loneos_sdss_color_dependence}
\end{figure}

However, the absolute magnitude
of RR Lyrae have been measured in $V$. We use photometry from SDSS
Data Release 3 (DR3) \citep{SDSS_DR3} to convert $M_{LONEOS\_F}$ to a
$V$-based system, in this case a synthetic SDSS-based magnitudes ($V_{SDSS}
= r + 0.44(g - r) - 0.02$ \citep{Ivezic2005}).  
All magnitudes have been corrected for Galactic
extinction using \citet{Schlegel}.  
The residual of $M_{LONEOS\_F}$ and $V_{SDSS}$ as a function
SDSS color, $g-r$, has a clear trend 
(Figure \ref{fig:loneos_sdss_color_dependence}).
Thus, we must restrict the range of SDSS color used to  
an empirical offset between $M_{LONEOS\_F}$ and $V_{SDSS}$. 
Stars in the DR3 with colors of RR Lyrae \citep{Ivezic2005} 
are matched to \loneosI stars ($0.99 < u - g < 1.28$, $-0.11 < g - r < 0.31$, 
$-0.13 < r -i < 0.2$, $-0.19 < i - z < 0.23$).  
A total of 40,180 SDSS DR3 stellar
objects in this color range were matched to \loneosI objects. Using
\loneosI timeseries, we reject stars with signs of variability. We see
that there is an offset, $M_{LONEOS\_F} - V_{SDSS} = -0.32$, with a
scatter of $0.13$. All subsequent \loneosI photometry in this work is
tied to the $V_{SDSS}$ system via this offset.  The final
transformation is:
\begin{equation}
M_{LONEOS \ V} = [M_{LONEOS \ F} - A_{GSC F} \times E(B-V)] + 0.32),
\end{equation}
where $A_{GSC F} = 2.65$ is the extinction coefficient and E(B-V) is
the reddening. This transformation is only valid for stars with 
colors similar to those of RR Lyrae stars.

\begin{figure}[t]
\epsscale{1.0}
\centering
\ifsubmode
\plotone{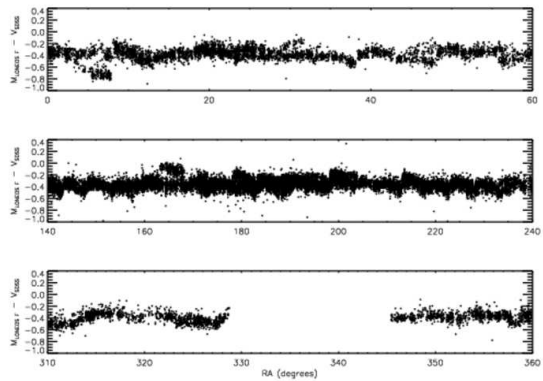}
\fi
\caption{The residual between LONEOS magnitudes (tied to GSC2) 
and SDSS as a function of RA for equatorial region. One can see 
the systematic difference between different POSS plates.}
\label{fig:gsc_sys_ra}
\end{figure}

The distribution of the residuals is approximately Gaussian in the
core, but there are long tails in the distribution, most likely due to
plate-to-plate systematic errors in the GSC photographic sky
survey. Figure \ref{fig:gsc_sys_ra} shows the residual of
$M_{LONEOS\_F}$ and $V_{SDSS}$ as a function of Right Ascension. One
can clearly see systematic errors as a function of RA.  These are on
the scale of Schmidt plate size (6 degrees). However, these large
errors occur over only a small fraction of the sky. 0.98\% of the stars
have residuals greater than  3 $\sigma$ from the mean 
compared to 0.3\% for a Gaussian distribution. 
Since the survey area which we will analyze 
is large, these errors do not affect the statistical properties of 
our RR Lyrae sample.  

\section{\loneosI {\it ab}-type RR Lyrae Star Catalog}
In this section, we will describe the selection of RR Lyrae using
\loneosI data, and present the \loneosI RR Lyrae star catalog. This
currently represents the largest single sample of halo RR Lyrae
stars. In addition, we discuss the detection efficiency of our
selection process and the completeness of our sample.

\subsection{Selection of \loneosI RR Lyrae Stars}
\begin{figure}[t]
\epsscale{0.9}
\centering
\ifsubmode
\plotone{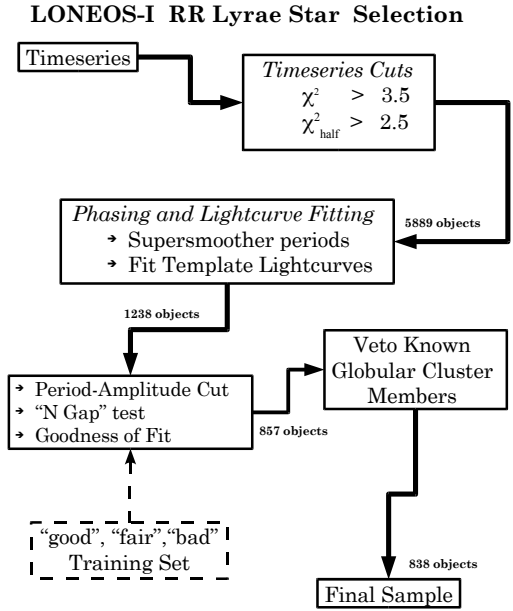}
\fi
\caption[\loneosI RR Lyrae Selection Flow Chart]{\loneosI RR Lyrae 
flow chart indicate the selection process and criteria used in the analysis
of this paper.  }
\label{fig:rr_flow_chart}
\end{figure}

The selection of the RR Lyrae was solely based on \loneosI temporal
photometric data.  No color information was used. We restrict our
search to fields with more than 28 good quality images. This amounts 
to 1430 $deg^{2}$ of coverage. The median number of images is 31.  
We impose a set of variability criteria which first 
identifies short-term variable stars, and then
analyzes the morphology of the phased light curves. We limit our
search to {\it ab}-type RR Lyrae, since it is difficult to detect
efficiently the {\it c}-type RR Lyrae given our temporal sampling.  In
order to accurately determine our detection efficiency, we did not
directly rely on visual inspection to produce our sample, but used a
series of cuts which aim to maximize the detection of RR Lyrae in an
objective manner. Below, we outline our selection criteria (see Figure
\ref{fig:rr_flow_chart}).

\begin{figure}[t]
\epsscale{1.0}
\centering
\ifsubmode
\plotone{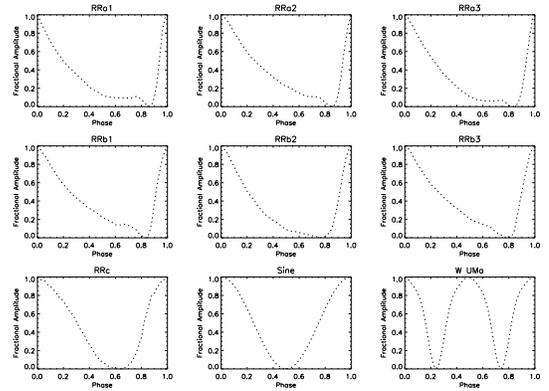}
\fi
\caption[RR Lyrae Templates]{The best four periods from the supersmoother 
are fit to these nine templates. The mean magnitude, phase and amplitude are 
free parameters in the fit. These nine templates are taken from \citet{Layden1998}. 
The first six are {\it ab}-type RR Lyrae which were obtained from the 
averaging of high-quality light curves. The last three include a {\it c}-type 
RR Lyrae, a sine wave, and W UMa contact binary. }
\label{fig:rr_templates}
\end{figure}

To reject non-variables and/or low amplitude variables, we made two
variability cuts on the timeseries:
\begin{itemize}
\item We first looked at the spread of the measured magnitudes about
        the average magnitude, normalized by the photometric error,
        which is a reduced $\Chisqrm$. We define this statistic as:
        $\Chisqrm = \frac{1}{N-1} \sum_{i=1}^{N} \frac{(m_{i} - \bar
        m)^{2}}{\sigma_{i}^{2}}$, where N is the number is
        observations, $m_{i}$ is the magnitude of the $i^{th}$
        observation, $\bar m$ is the mean magnitude, and $\sigma_{i}$
        is the photometric error of the $i^{th}$ observation.
        $\Chisqrm$ gives a measure of the variability over the entire
        timeseries.  We rejected stars with $\Chisqrm < 3.5$. This
        predominantly rejected small amplitude variables.

\item We divided the timeseries into two segments, each containing
        half the data points, and calculated $\Chisqrm$ for both
        halves. We required $\Chisqrm > 2.5$ for both halves of the
        timeseries. This rejects objects that are aperiodic or that
        have inflated $\Chisqrm$ from occasional outliers.
\end{itemize}
A total of 5889 objects satisfied these cuts. In order to identify RR
Lyrae stars in this sample, we search for a periodic signal in these
timeseries. In addition, we must classify Period-folded light
curves. However, the \loneosI sampling is not sufficient to use
sophisticated classification methods, such as Fourier decomposition or
machine learning methods.  We use template light curves to distinguish
RR Lyrae from other common light curves, such as sinusoidal variables
and eclipsing binaries.  We use the nine templates from
\cite{Layden1998}, based on high quality light curves of bright RR
Lyrae stars which have been averaged.  Six templates are {\it ab}-type
RR Lyrae stars. In addition, there are single templates for {\it
c}-type RR Lyrae, for a sinusoid, and for a W UMa-type contact binary
(Figure \ref{fig:rr_templates}). There are a sizable number of
contact binaries in the \loneosI database.  The fundamental mode RR
Lyrae {\it ab}-type can be broken down into two different categories,
type {\it a} and {\it b} (i.e., ``Bailey'' type).  Bailey type {\it a} objects have
flatter bottoms, and the Bailey type {\it b} have a sawtooth shape.  The
first six templates represent typical RR Lyrae light curve shapes for
type {\it a} and {\it b} with three different rise times (i.e., the phase interval
between maximum and minimum light).

We used a supersmoother routine \citep{Reimann1994} to obtain a set of
four best periods.  
The supersmoother is a variable span
smoother, where the best span is chosen using a cross-validation
technique. It was developed for the MACHO experiment
\citep{MACHO1995}.  We have tested other standard routines, such as
Phase Dispersion Minimization \citep{Stellingwerf1978} and
Lomb-Scargle periodogram \citep{Scargle1982}; these alternative
algorithms provide comparable results to the supersmoother.  The
supersmoother makes no assumptions regarding the morphology of the
light curve.  As a consequence, the supersmoother's best period may
not represent a physically realistic light curve. In addition, we have
found that light curves with outlying data points adversely affect the
supersmoother's figure of merit for smoothness.

We limited the period search to range between 0.25 and 0.9 days. Our
results were not affected if the search is extended to 2.0 days.  In
order to break degeneracies between the four best periods returned, we
used a non-linear $\Chisqrm$ minimization method to fit the phased
data to our set of template light curves (Figure
\ref{fig:rr_templates}). The magnitude offset, phase and amplitude
were free parameters in the template fitting. We required that the fit
amplitudes be greater than $0.3$ magnitudes. Each of the four periods
was fit separately. The best period was selected by the quality of the
template fit.

Objects were tentatively classified as {\it ab}-type RR Lyrae if
convergence of the non-linear minimization was satisfied for the fits
to one of the six RR{\it ab} templates. We defer any template light
curve goodness-of-fit cuts until later.  Objects with best fits to the
three non-RR{\it ab} template light curves were discarded.  A total of
1238 objects were classified as {\it ab}-type RR Lyrae.  These objects
were visually inspected and classified into three groups: ``good'',
``fair'', and ``bad''.  The criteria for the classifications included:
goodness-of-fit to template light curve, and uniformity of data across
phase. It is fairly easy to classify objects in the ``good'' and
``bad'' categories. The ``fair'' group includes objects where there
was some ambiguity in the classification.  These classifications were
used as a training sample for further selection criteria. The aim was
to determine a totally automated set of selection criteria to maximize
the selection of real RR{\it ab} while minimizing contamination. This
will facilitate the calculation of our detection efficiency.  We
developed the following criteria for quality of the light curves.
First, we required that the fitted periods and amplitudes reside in
the region of period-amplitude space where RR{\it ab} are known to
lie. This region was defined to contain the bulk of our ``good''
sample, and was large enough to encompass the range of metallicities
and Oosterhoff groups associated with halo RR Lyrae stars. We are
guided by the MACHO RR Lyrae period-amplitude distribution for the
Galactic Bulge \citep{MACHO_Bulge_RR}, which contains a population of
RR Lyrae stars with a wide range of metallicities. The region in
period-amplitude space was defined as:
\begin{equation}
A < 1.5 \ {\rm magnitudes},
\end{equation}
\begin{equation}
A > 0.3 \ {\rm magnitudes},
\end{equation}
\begin{equation}
A > -5 \ log(P) - 1 \ {\rm magnitudes},
\end{equation}
\begin{equation}
A < -5 \ log(P) + 0.3 \ {\rm magnitudes},
\end{equation}
where periods are measured in days. Using the well-sampled light
curves from Northern Sky Variability Survey (NSVS),
\cite{Kinemuchi2006} found a population of {\it ab}-type RR Lyrae with
shorter periods than the OoI group.  Using light curve Fourier
decomposition techniques, \cite{Kinemuchi2006} determined that this
population was more metal-rich than the rest of the ROTSE-I
sample. This population was identified as a metal-rich thick disk RR
Lyrae star population.  Our period-amplitude cut largely excludes this
metal-rich thick disk RR Lyrae star population.  A metal-weak thick
disk RR Lyrae star population also likely exists, but this population
smears with the OoI in the period-amplitude diagram.  Thus, we are
confident that out our RR Lyrae star sample will consist predominantly
of halo objects. Of course, kinematic information is required to be
certain of population membership.  Finally, an OoII component may
extend to smaller amplitudes than an OoI component.  The amplitude cut
should optimally be extended to smaller amplitudes at longer periods,
but our detection efficiency below 0.3 magnitudes is relatively low at
the faint end.  While this cut may preferentially remove OoII RR Lyrae
stars, the \loneosI dataset cannot efficiently probe this region
period-amplitude space. 

Using our visual classifications (i.e., ``good'', ``fair'', and
``bad''), we determined the thresholds for two additional cuts:
uniformity and smoothness.  These cuts were performed simultaneously,
and their thresholds were set to minimize the number of ``bad''
objects and maximize the number of ``good'' objects.  First, we
required that the light curves have a degree of uniformity. The phased
data points should be distributed from zero to one as randomly and
uniformly as possible. In order to measure this, we divided the phased
light curve into 10 equally spaced bins, and binned the data points.
We required that at least 7 bins have at least one measurement.  Thus,
at most 30\% of the phase is allowed to contain no data points. We
referred to this as the ``N gap'' test.  Secondly, we examined the
goodness of fit of the RR Lyrae star template relative to the fit to a
constant magnitude. We define this statistic as: $\Delta \Chisqrm =
(\Chisqrm_{constant} - \Chisqrm_{RR})/\Chisqrm_{RR}$, where we use a
reduced $\Chisqrm$ for all the fits. This statistic measures the
scatter about the six RR{\it ab} templates, and thus smoothness of the
period-folded data.  For this sample, we used $\Delta \Chisqrm > 5$.

\begin{figure}[t]
\epsscale{1.0}
\centering
\ifsubmode
\plotone{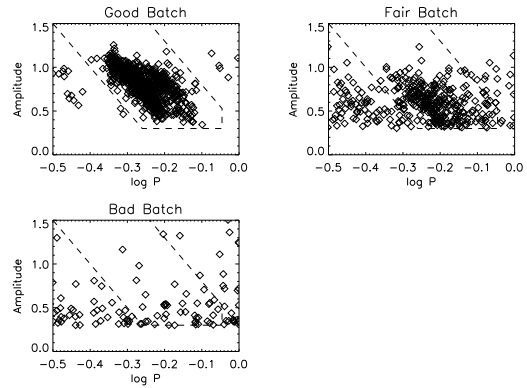}
\fi
\caption[Period-Amplitude Plots after template fitting]{These three 
period-amplitude distributions show the objects which were successfully 
fit to an RR Lyrae template light curve. The plots show the amplitude
(in \loneosI instrumental magnitudes) versus the logarithm of the period 
measured in days. They have been manually classified into three categories,
``good'', ``fair'', ``bad''. The ``good'' and ``fair'' distributions 
are consistent with being RR Lyrae with some scatter, while the 
``bad'' category is randomly distributed.
The dashed lines indicates the region of period-amplitude space where 
RR Lyrae stars are expected to lie.
This region is large enough to
encompass a wide range of metallicities and Oosterhoff types. 
}
\label{fig:BEFORE_per_amp_cuts}
\end{figure}

\begin{figure}[t]
\epsscale{1.0}
\centering
\ifsubmode
\plotone{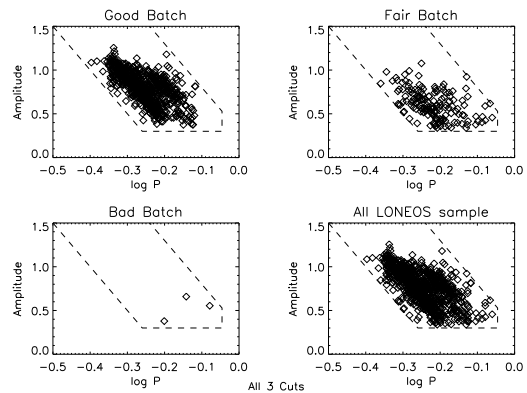}
\fi
\caption[Period-Amplitude Plots after 3 cuts]{We show the resulting objects 
after we have applied the period-amplitude cut and
two additional light curve quality cuts described in the text. 
The lower-right panel shows the final RR Lyrae sample.}
\label{fig:AFTER_per_amp_cuts}
\end{figure}

\begin{figure}[t]
\epsscale{1.0}
\ifsubmode
\plottwo{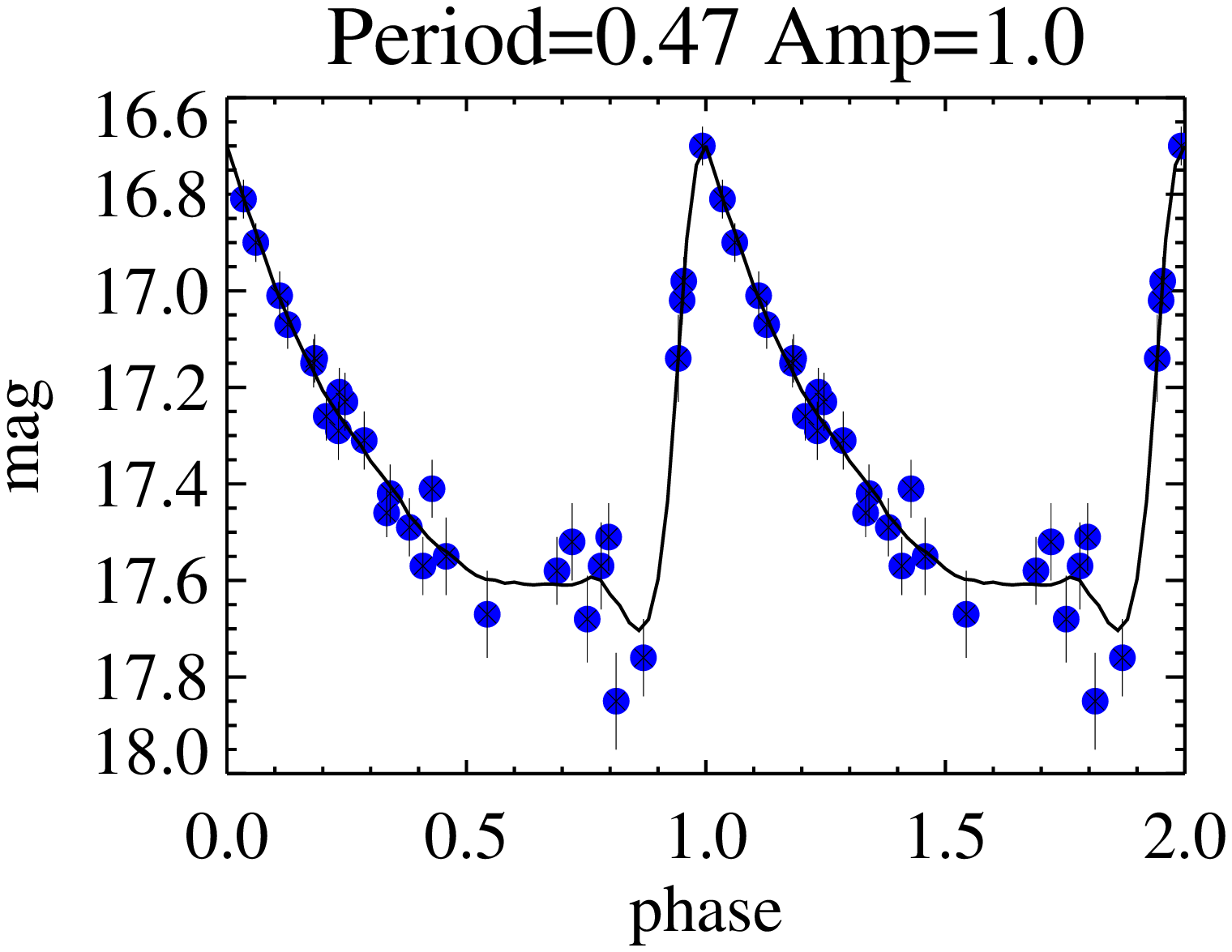}{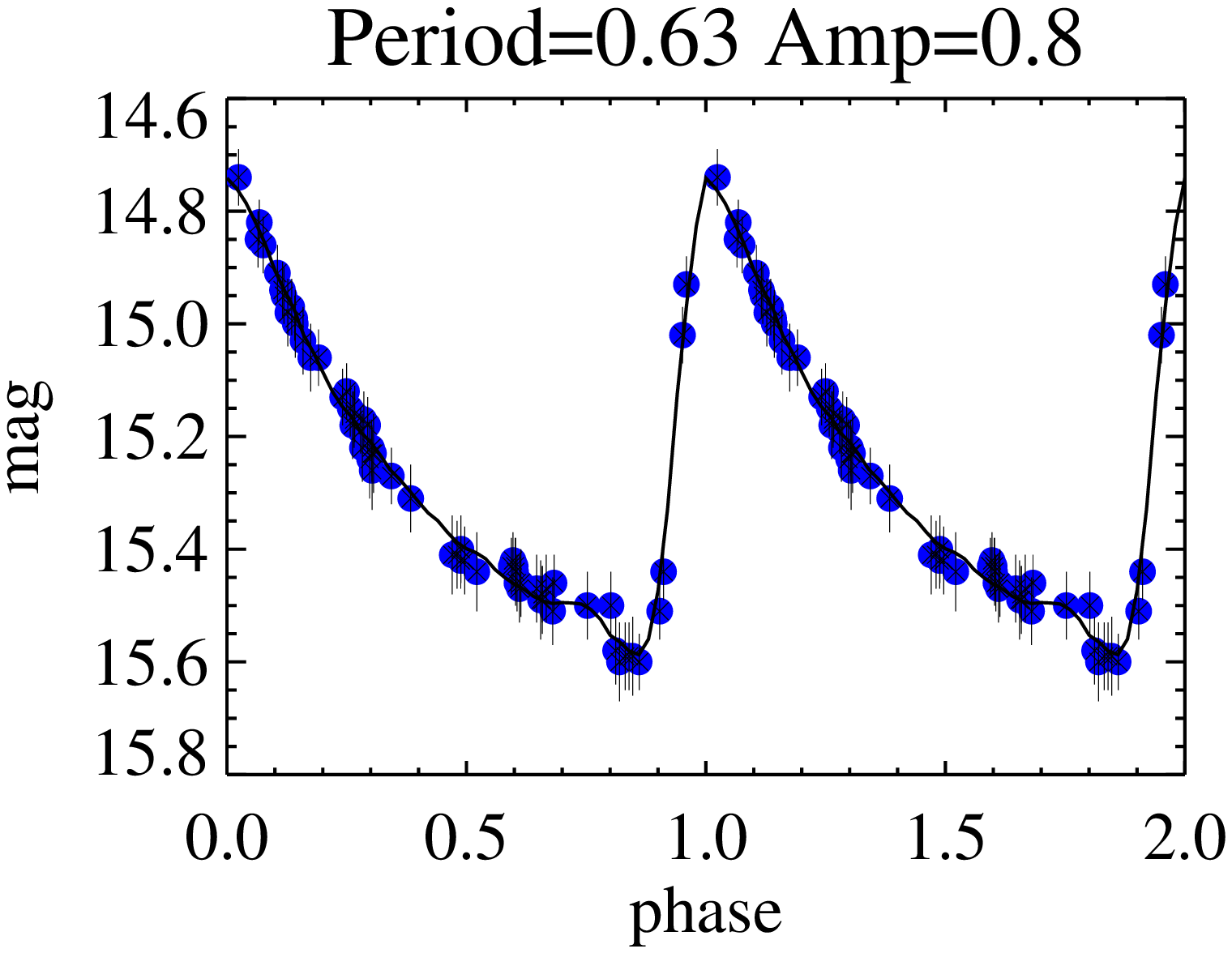}
\fi
\caption[Sample of \loneosI RR Lyrae period-folded light curves]{
Two examples of phased RR Lyrae light curves. The lightcurves are shown over two full phases.
\label{fig:lightcurve2}}
\end{figure}

Together, these three cuts reject 8\%, 60\%, 97\% from the ``good'',
``fair'', and ``bad'' samples, respectively.
Figures \ref{fig:BEFORE_per_amp_cuts} and \ref{fig:AFTER_per_amp_cuts}
show the results of the final three cuts.  The sample presented in
this paper are restricted to mean magnitudes ranging from 13.0 to
17.5. We will discuss the completeness of this sample in the following
sections.  We also removed known RR Lyrae that are members of globular
clusters and RR Lyrae within the tidal radius of the globular clusters
in our survey volume, a total of 19 objects.  Our final sample
consists of 838 {\it ab}-type RR Lyrae, which are solely based on
temporal photometric data. Table \ref{rr_table} lists the properties
of this sample.  Two sample light curves are shown in Figure
\ref{fig:lightcurve2}. To date, this represents the largest single
sample of period-folded RR Lyrae stars in the stellar halo.
\tabletypesize{\scriptsize}
\begin{deluxetable}{ccccc}
\setlength{\tabcolsep}{0.006in} 
\tablecolumns{5}
\tablewidth{0pc}
\tablecaption{Properties of 838 \loneosI RR Lyrae Sample
\label{rr_table}}
\tablehead{
\colhead{RA} & \colhead{DEC} & \colhead{ $<M_{LONEOS \ V}>$} & \colhead{Period} & \colhead{Amplitude}  }
\startdata
231.923019 &     -22.413485  &    14.48 &  0.606328     &   0.890   \\
234.829575 &      -22.006363 &     16.45 &  0.644697   &     0.734   \\
303.111725 &     -23.159113  &    15.18  & 0.608629   &     0.812   \\
302.516479 &     -22.789234  &    15.62 &  0.670238  &      0.705   \\
\vdots     &         \vdots     &     \vdots &     \vdots    &    \vdots \\
259.791656 &       45.140347    &    15.25   &     0.468290  &      1.083        \\    
\enddata
\tablecomments{
This table list a few properties of the
838 \loneosI RR Lyrae sample. The coordinates (RA,DEC) are given in decimal degrees 
in the epoch of J2000.0. The mean magnitudes ($<M_{LONEOS \ V}>$) are tied to the $V_{SDSS}$ system.
The periods are given in days. Finally, the amplitudes are measured from minimum to 
maximum brightness as defined by the RR Lyrae template that best fit the lightcurve.
[The complete version of this table is in the electronic edition of
the Journal.  The printed edition contains only a sample.]
}
\end{deluxetable}

\begin{figure}[t]
\epsscale{1.0}
\centering
\ifsubmode
\plotone{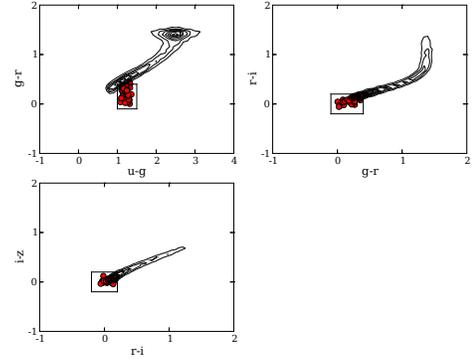}
\fi
\caption[SDSS color-color diagrams for \loneosI RR Lyrae]{These color-color 
diagrams are constructed from SDSS photometry. The contours indicate
the location of the stellar locus in each of the 2D slices through the
SDSS color-color space.  The rectangular areas are the RR Lyrae color
boundaries are from \citet{Ivezic2005}.  The red points are \loneosI
RR Lyrae (78) with SDSS DR3 photometry. There are only two objects
which do not pass the $g - r$ cut, but are consistent with RR Lyrae
stars in the three other colors.  Thus, virtually all of these objects
have colors consistent with RR Lyrae stars.
}
\label{fig:loneos_sdss}
\end{figure}

As a simple test of the contamination of our RR Lyrae sample, we
explored the colors of a subset of these objects. RR Lyrae should be
be identifiable on the basis of Sloan Digital Sky Survey (SDSS)
photometry alone \citep{KK_thesis, KK98}.  78 of the \loneosI RR Lyrae have clean
SDSS Data Release 3 photometry \citep{SDSS_DR3}.  Figure
\ref{fig:loneos_sdss} shows the color-color diagrams constructed from
SDSS DR3 photometry. RR Lyrae stars are contained in a small region of
SDSS color-color space \citep{Ivezic2005} as shown by the rectangular
areas outlined in the figure.  The contours indicate the location of
the stellar locus in each of the 2D slices through the SDSS
color-color space. The red points are \loneosI RR Lyrae stars (78)
with SDSS DR3 photometry. There are only two objects which do not pass
the $g - r$ cut, but are consistent with RR Lyrae stars in the three
other colors.  Thus, virtually all of these objects have colors
consistent with RR Lyrae stars. We conclude that our RR Lyrae sample
has a low degree of contamination.

\begin{figure}[t]
\epsscale{1.0}
\centering
\ifsubmode
\plotone{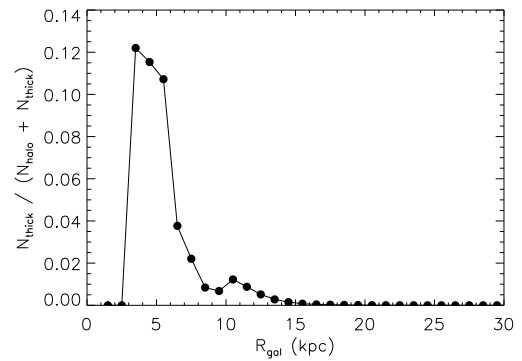}
\fi
\caption[Thick Disk Contamination]{This figure shows the predicted fraction of thick 
disk stars relative to halo stars which is expected
in the \loneosI survey volume as a function of Galactocentric Radius. }
\label{fig:thick_disk_fraction}
\end{figure}

The survey contains fields with Galactic latitude as low as 10 degrees. 
Thus, there is a possibility that RR Lyrae from the
thick disk have contaminated our sample.  Our period-amplitude cut
likely has removed most of the thick disk RR Lyrae, but let us
investigate how much thick disk contamination our sample might have
using Galactic models.  We use a double exponential to model the thick
disk with a scale height of 0.7 kpc and scale length of 3.0 kpc
\citep{Layden1995}.  These scales are based on measurements using RR
Lyrae stars.  Our halo is modeled as a power law with an index of
$-2.43$, which is the value we found using the \loneosI RR Lyrae (See
the next section for details). We assume that the thick disk is 10
times more populous than the stellar halo in the Solar Neighborhood,
and the measured stellar number density is $0.1 pc^{-3}$ \citep{NLDS}.
We also assume that the frequencies of RR Lyrae stars in the thick
disk and halo are identical.  This is a conservative assumption since
thick disk RR Lyrae stars frequencies will be suppressed since thick
disk RR Lyrae stars are more metal-rich.  Given these parameters, we
calculate the anticipated fraction of thick disk stars relative to
halo stars in the \loneosI survey volume. Figure
\ref{fig:thick_disk_fraction} shows this fraction as a function of
Galactocentric Radius. This figure shows that the fraction of thick
disk stars to halo stars is about 12\% even at the smallest radii,
where the largest contamination is expected.  This low fraction of
thick disk stars and our period-amplitude cut give us confidence that
the contamination of thick disk stars in our sample is negligible.
 
\begin{figure}[t]
\epsscale{1.0}
\centering
\ifsubmode
\plotone{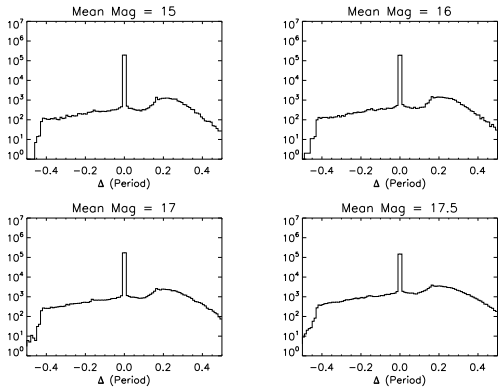}
\fi
\caption[Period Error]{We have simulated RR Lyrae 
light curves with a range of periods and examined the periods recovered using
our phasing and light curving fitting described in Section 3.1.
The histograms show the difference between the input and recovered periods for four mean 
magnitudes. The fraction contained within $\pm 0.01$ days
are: 0.92, 0.90, 0.81, 0.70, for mean magnitudes of 15,16,17, and 17.5, respectively.  }
\label{fig:period_errors}
\end{figure}

In order to investigate how well we recover various RR Lyrae light
curve parameters, we simulated RR Lyrae light curves with known
parameters and examined the recovered parameters using 
the RR Lyrae period and light curve fitting procedure described above.
We used the actual cadence for each subfield to generate simulated 
time series of RR Lyrae stars using template RR Lyrae light 
curves from Layden (1998) with various mean magnitudes. 
We used a known RR Lyrae period-amplitude distribution
\citep{Smith_book} to generate the simulated time series with random phase.
We assigned photometric error bars and smear the simulated light curves
based on the magnitude dependence of the \loneosI photometric errors.
In Figure \ref{fig:period_errors}, the residuals of the input and recovered 
periods are shown.
Periods are recovered with good accuracy.  The fraction of simulated RR Lyrae
contained within $\pm 0.01$ days are: 0.92, 0.90, 0.81, 0.70, for mean
magnitudes of 15, 16, 17, and 17.5, respectively.  We will investigate
the possibility of misidentified periods in the next section when we
discuss the Oosterhoff components.

\begin{figure}[t]
\epsscale{1.0}
\centering
\ifsubmode
\plotone{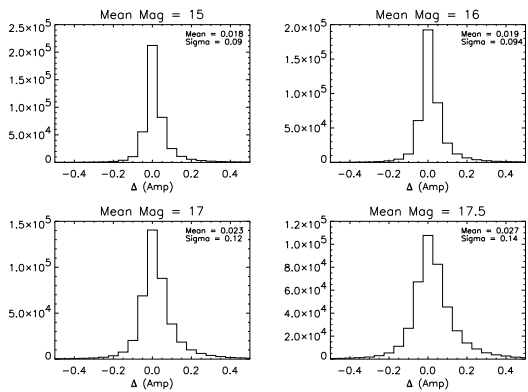}
\fi
\caption[Amplitude Error]{These histograms show the the residual between the  
input and recovered amplitudes from simulated RR Lyrae light curves of various mean
magnitudes. The residual increases with mean magnitude. In addition, 
the distributions are skewed; the recovered amplitudes tend to be underestimated.
}
\label{fig:amp_errors}
\end{figure}

Similarly, Figure \ref{fig:amp_errors} shows the residuals 
of the input and the 
recovered amplitude. The residual increases with mean magnitude. 
In addition, the distributions are skewed; the
recovered amplitudes tend to be underestimated. This is a consequence
of the asymmetric shape of {\it ab}-type RR Lyrae light curves, and
the sparseness of our sampling. However, we find that the
underestimation of  amplitudes does not depend on the input
period. Thus, underestimating the amplitude will only have the effect
of shifting RR Lyrae stars downward in a period-amplitude diagram.  We
will discuss this more in the next section.

\subsection{Completeness}
In this section, we estimate our efficiency in detecting RR Lyrae
stars (i.e., the completeness of the sample). The completeness depends
on a number of RR Lyrae parameters, but most strongly on their
magnitude and amplitude. One of our scientific goals is to measure the
number density of RR Lyrae stars as a function of position in the
Galaxy. This requires that we estimate the detection efficiency as a
function of Galactic position.

We divided our completeness analysis into photometric and temporal
parts.  In order to estimate the photometric efficiency, we take a
clean, unblended, and dereddened sample of stellar objects from SDSS
DR3, which are contained in the \loneosI footprint. Since the SDSS
detection limits are at least 2 magnitudes deeper than LONEOS, we can
use SDSS to understand the degree to which \loneosI is photometrically
complete.  Using this sample, we performed a proximity query using our
\loneosI database.  For each SDSS star, we search in the \loneosI
database for objects within 1.8 arcsecs.
A total of 58,591 SDSS stars are queried in the \loneosI database.
Next, we count the number of detections and non-detections of the
object in every \loneosI image. A measurement is considered a
non-detection if any one of the following criteria are true:
\begin{itemize}
\item Signal-to-noise ratio is below threshold ($5 \sigma$)
\item Measurements are flagged by photometric pipeline as bad (e.g.,
blended, pixel defects, non-stellar shape, cosmic ray hit)
\item More than one \loneosI object is detected, which indicates
problems with time series generation.
\end{itemize}
These criteria are also used for the later analysis.

\begin{figure}
\epsscale{1.0}
\centering
\ifsubmode
\plotone{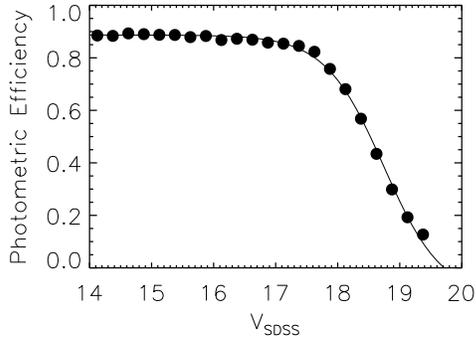}
\fi
\caption[\loneosI photometric efficiency]{We have taken a clean sample of 
unblended SDSS DR3 stars and queried the \loneosI database for a detection 
in each image in the database. This figure shows the overall
photometric completeness (image-by-image) assuming SDSS is complete. 
Note that non-detection also include measurements that were flagged as 
being problematic.}
\label{fig:dr1_phot_complete}
\end{figure}

Thus, this comparison with SDSS not only gives us an estimate of the
photometric completeness based on signal-to-noise, but also an
estimate of the efficiency of the photometric pipeline. The last two
non-detection criteria are generally independent of magnitude. Thus,
these will contribute to an overall reduction across all magnitudes.
This is why the efficiency converges to only $\sim 90\%$ in the bright
magnitude range. Figure \ref{fig:dr1_phot_complete} shows the \loneosI
photometric completeness. We use the following parameterization for
photometric efficiency shown in Figure \ref{fig:dr1_phot_complete}:
\begin{equation}
f(V_{SDSS}) = \frac{1}{1 + e^{\frac{V_{SDSS} - 18.74}{0.47}}} - 0.11.
\end{equation}
In addition, we did not find a significant dependence on the
photometric completeness curves with position on the sky. Thus, we use
this as our universal photometric completeness function.

\begin{figure}
\epsscale{1.0}
\centering
\ifsubmode
\plotone{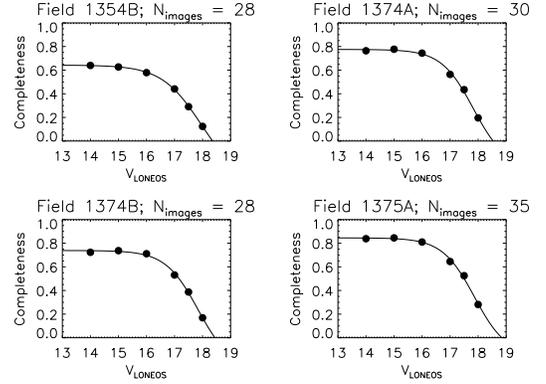}
\fi
\caption[Sample of Completeness Curves for Four Subfields.]{
The RR Lyrae star completeness (photometric and temporal) for four
fields with different number of images are shown. Note that even 
though fields 1354B and 1374B have identical numbers of
images, the completeness curves are different. This is caused by the different 
cadences of these two fields.
These curves are used to correct the effective volume in the density measurements.}
\label{fig:sample_completeness_curves}
\end{figure}

\begin{figure}
\epsscale{1.0}
\centering
\ifsubmode
\plotone{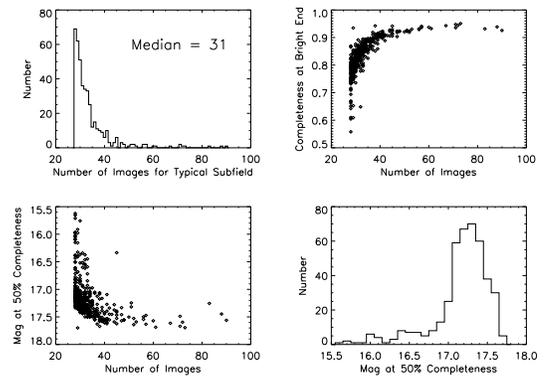}
\fi
\caption[Completeness Properties of \loneosI Subfields.]{
This figure shows various parameters and correlations related to the RR Lyrae star 
completeness (photometric and temporal). First, the top left plot is a
histogram of the number of images for the typical subfields. The median
number of images is 31. 
The top right scatter plot shows the dependence of the magnitude 
at the bright end (1-Z in Equation \ref{equ:completeness}) 
on the number of images. This gives of sense of the completeness 
for bright stars, where signal-to-noise ratio is high.
On the other hand, the bottom left scatter plot shows the dependence 
of the magnitude at 50\% completeness on the number of images, 
which gives a sense of the completeness towards
the faint end. Finally, the bottom right histogram shows the 
distribution of the magnitude at 50\% completeness.  
}
\label{fig:temporal_complete}
\end{figure}

Using this parameterized photometric completeness, we performed the
temporal completeness analysis. 
We generated simulated RR Lyrae times series in the same manner  
as described in Section 3.1, except
that we incorporated the parameterized photometric completeness, 
which provided a detection probability as function of magnitude. 
We randomly rejected individual time series measurements 
based on this detection probability.
Given the large number of subfields ($\sim
15,000$), we performed this analysis on one ``typical'' subfield per
field, which had the median number of images; the number of images for
subfields in a given field does not vary much. Thus, we perform this
analysis on 421 ``typical'' subfields.  We generated 1000 simulated time
series for each ``typical'' subfield and for each magnitude. 
All the simulated RR Lyrae times series were analyzed using 
the same selection criteria  (see Figure \ref{fig:rr_flow_chart}) 
as those used to generate the real \loneosI RR Lyrae catalog.
The completeness functions were generated based on the percentage of 
simulated time series which passed our RR Lyrae selection criteria.
The completeness functions as a function of magnitude (M) for the
``typical'' subfields were parametrized by the following function:
\begin{equation}
f'(M_{LONEOS \ V}) = \frac{1}{1 + e^{\frac{M_{LONEOS \ V} - \mu}{T}}}
- Z, \label{equ:completeness}
\end{equation}
where $\mu$ is the characteristic magnitude cutoff, $T$ is the width
of the cutoff, and $Z$ is a horizontal offset. Figure
\ref{fig:sample_completeness_curves} shows the detection efficiency as
a function of magnitude for four fields. All magnitudes have been
corrected for Galactic extinction using \cite{Schlegel}.  These
efficiencies are used when we calculate the survey volume and
corrected RR Lyrae star number densities in the following section. 
The completeness function for the ``typical'' subfield is used 
for all subfields of the same field. A histogram of the number of 
images in the ``typical'' subfields is shown in the top left plot of
Figure  \ref{fig:temporal_complete}. The median number of 
images is 31. In order to give a sense of 
the dependence of the completeness on  
the  number of images, Figure \ref{fig:temporal_complete} shows the 
correlations of the magnitudes at 50\% completeness 
and the completeness at the bright end (1-Z in Equation
\ref{equ:completeness}) with the number of images in 
the top right and bottom left plots, respectively.
The completeness is generally correlated with the number of images.


\section{Analysis}
In this section, we explore the bulk properties of the \loneosI RR
Lyrae star sample that was described in the previous section.  First,
we present the spatial distribution of the entire sample. Next, we
present the period-amplitude distribution of the sample.  Finally, we
explore the spatial distributions of the different populations that
emerge from the period-amplitude distribution.

\subsection{Radial Density Distribution}
As described in previous sections, we tie our absolute photometry to
synthetic SDSS-based magnitudes ($V_{SDSS}$), $M_{LONEOS \ V}$.  This
allows for a straightforward calculation of distances from the
intensity-weighted mean magnitudes.  Distances are calculated from the
following formula:
\begin{equation}
d = 10 ^{\frac{ <M_{LONEOS \ V}> - M_{V} - 10 }{5}} \ kpc,
\end{equation}
We will use RR Lyrae absolute magnitudes of $M_{V} = +0.71 \pm 0.12$
\citep{Layden96} derived from statistical parallax analysis for a
sample with $\langle[Fe/H]\rangle = -1.61$.
Assuming that the Sun is at $-7.8$ kpc \citep{Carney_Galactic_Center}
from the galactic center along the y-axis ($x_{gal},y_{gal},z_{gal} =
(0, -7.8, 0)$ kpc), we calculate the galactocentric 3-D coordinates of
each RR Lyrae star.

\begin{figure}
\epsscale{1.0}
\centering
\ifsubmode
\plotone{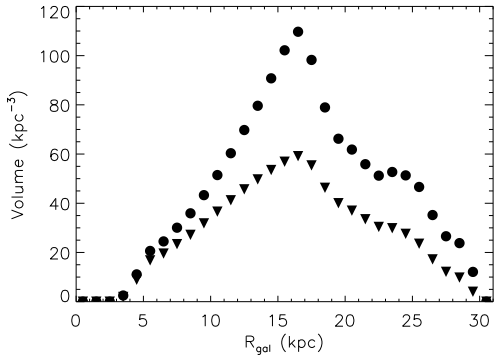}
\fi
\caption[Raw and Detection Efficiency-Corrected Survey Volume]{This figure 
shows the raw (circles) and the detection
efficiency-corrected (triangles) survey volumes. Note that the difference 
between the two curves is not monotonically increasing
because there are lines-of-sight which are pointed towards the 
Galactic Center and also the anti-Center which contribute at
different ranges of Galactocentric radius. If the volume calculation 
assumes that the Sun is at the center of the Galaxy, the
difference between these two curves does indeed increase monotonically. }
\label{fig:volume}
\end{figure}

We measure the spherically symmetric radial density distribution by
counting the number of RR Lyrae stars in galactocentric radial bins
and calculating the survey volume contained in each bin.  We use 1 kpc
bins for all the density distributions. This provides us with a
measure of the radial density distribution averaged along the survey
lines-of-sight.  Using the boundaries of the survey we numerically
integrate the volume contained in each radial bin, taking into account
detection efficiency at each position in the Monte Carlo integration.
Figure \ref{fig:volume} shows the survey effective volume
distribution.

\begin{figure}
\epsscale{1.0}
\centering
\ifsubmode
\plotone{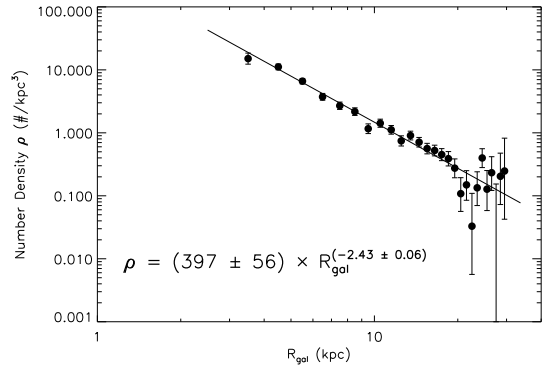}
\fi
\caption[\loneosI RR Lyrae Radial Distribution ($M_{V} = +0.71$)]{The radial 
density distribution of 838 {\it ab}-type
RR Lyrae with a mean magnitude range
from 13.0 to 17.5, using $M_{V} = +0.71$. We do not find any 
statistically significant
deviations from a smooth stellar halo. Our survey volume does not 
intersect known
locations of the tidal debris from the Sagittarius dwarf galaxy.  }
\label{fig:radial_all}
\end{figure}

Assuming a spherically symmetric distribution, Figure
\ref{fig:radial_all} shows the radial profile for our sample with mean
magnitudes in the range [13,17.5] with $M_{V} = +0.71$. The one-sigma,
asymmetric Poisson error bars are assigned and calculated using the
method of \citet{Gehrels}.  Using non-linear least squares, we fit
power laws to the radial density distributions:
\begin{equation}
\rho(R_{gal}) = (397 \pm 56) \times R_{gal}^{(-2.43 \pm 0.06)} \
(kpc^{-3}).
\end{equation}
with \Chisqr per degree of freedom of 0.91 .  All fitted 
parameter estimates are calculated using a Monte Carlo 
bootstrap technique \citep{NR}.
Table \ref{table:other_surveys} summaries results from other surveys.  This
result is within $1-\sigma$ of the QUEST-I (\cite{Vivas2006}, $M_{V} =
+0.55$, $R_{SUN} = 8.0$ kpc) and the SDSS (\cite{Ivezic00}, $M_{V} =
+0.7$, $R_{SUN} = 8.0$ kpc) results, assuming a spherical halo
distribution.  The power law exponent is slightly sensitive to the
distance to the Galactic Center.  If $R_{SUN} = 8.5$ kpc, then
exponent is reduced to $-2.50 \pm 0.05$.  As shown in Table
\ref{table:other_surveys}, the measured distance to the Galactic
Center has been decreasing with time. This means that earlier
measurements will tend to have steeper profiles than our measurement.
\tabletypesize{\tiny}
\begin{deluxetable*}{lcccccc}
\setlength{\tabcolsep}{0.02in} 
\tablecaption{Halo RR Lyrae Survey Comparisons
\label{table:other_surveys}}
\tablehead{
\colhead{Survey} & \colhead{Tracer} & \colhead{Exponent} &
\colhead{Flattening Ratio (q)} & \colhead{Objects} & \colhead{$R_{0}$ (kpc)} & \colhead{Comment}\\ }
\startdata
\loneosI (All) & RR Lyrae &  $-2.43 \pm 0.06$  &  $q = 1$ & 838 & 7.6  &  $M_{V} = +0.71$ \\
\loneosI (All) & RR Lyrae &  $-3.15 \pm 0.07$  &  Variable q & 838 & 7.6  &  $M_{V} = +0.71$ \\
\loneosI (Oo I) & RR Lyrae &  $ -2.26 \pm 0.07 $  &  $q = 1$ & 619 & 7.6  &  $M_{V} = +0.71$ \\
\loneosI (Oo II) & RR Lyrae &  $ -2.88 \pm 0.11 $  &  $q = 1$ & 219 & 7.6  & $M_{V} = +0.71$  \\
QUEST-I (\cite{Vivas2006}) &   RR Lyrae &  $-2.5 \pm 0.1$  &  $q = 1$ & 395 & 8.0 & $M_{V} = +0.55$  \\
QUEST-I (\cite{Vivas2006}) &   RR Lyrae &  $-3.1 \pm 0.1$  &  Variable q & 395 & 8.0 & $M_{V} = +0.55$ \\
SDSS (\cite{Ivezic00}) &   RR Lyrae &  $-2.7 \pm 0.2$  &  $q = 1$ & 148 & 8.0  & $M_{V} = +0.7$ \\
\cite{Chiba_Beers00} &    Halo ($\langle[Fe/H]\rangle \leq -1.8$) &  $-3.55 \pm 0.13$  &  Variable q ($q_{0} \sim 0.65$)  & 413 & \nodata  &  Kinematics  \\
\cite{Chiba_Beers00} &    Halo ($-1.6 < \langle[Fe/H]\rangle \leq -1.0$)  &  $-3.47 \pm 0.18$  & Variable q ($q_{0} \sim 0.55$) & 331 & \nodata & Kinematics   \\
\cite{Wetterer96}    &   RR Lyrae &  $-3.0 \pm 0.08$  &  $q = 1$   & 42 & 7.6 &  $M_{V} = +0.74$  \\
\cite{Wetterer96}    &   RR Lyrae &  $-3.5 \pm 0.08$  &  Variable q   & 42 & 7.6 &  $M_{V} = +0.74$  \\
\cite{Preston91} & RR Lyrae &  $-3.2 \pm 0.1$ & Variable q  & 47 & 8.0  &  $M_{V} = +0.6$  \\
\cite{Saha85} & RR Lyrae & $-3.0 $ & $q = 1$  & 29 & 8.7 & P-L-A relation ($M_{B} = M_{bol} +0.34)$ \\
\cite{Zinn1985} & Globular Clusters & $-3.5$  & \nodata   & 121 & \nodata & \nodata  \\
\cite{Hawkins84} & RR Lyrae & $-3.1 \pm 0.2$ & $q < 0.9$  & 24     & 8.7  & $M_{B} = +0.9$   \\
\cite{Oort75} & RR Lyrae & $-3.0 $ & $q < 0.8$ & 1108   & 8.7 &  $M_{B} = +0.9$ \\
\cite{Kinman66}  & RR Lyrae & $-3.5 $ & q= 0.6 - 0.8 & 38 &  10.0 &   $M_{V} = +0.42$  \\
\enddata
\tablecomments{
We compare the \loneosI RR Lyrae survey results to those of other surveys using RR Lyrae stars and other tracers.
}
\end{deluxetable*}

\begin{figure}
\epsscale{1.0}
\centering
\ifsubmode
\plotone{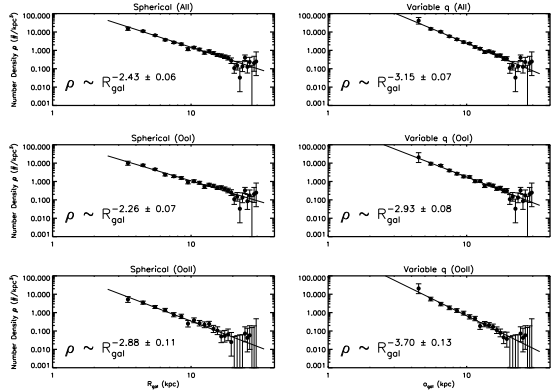}
\fi
\caption[Radial density profiles for spherical and variable flattening models]{
Radial density profiles for spherical and variable flattening models
are shown in the left and right columns, respectively. The variable
flattening model is that of \citet{Preston91}, where the flattening
ratio variable from 0.5 to 1.0 with increasing distance from the
Galactic Center.  The top row of plots are for the entire sample of RR
Lyrae stars. The second and third rows are for sub-samples of RR
Lyrae, classified as Oosterhoff I and II, respectively. The \loneosI
RR Lyrae data are fit by the spherical and variable flattening models
equally well. The fits are not statistically different.
}
\label{fig:spherical_and_variable_q_power_laws}
\end{figure}

The shapes of the stellar and dark matter halos should have encoded in
them some information of their formation process. In general, the
isodensity contours for the RR Lyrae number density can be
described by an ellipsoid:
\begin{equation}
\frac{x^{2}}{a^{2}} + \frac{y^{2}}{b^{2}} + \frac{z^{2}}{c^{2}} = 1.
\end{equation}
Let us consider the case where $a=b$ (i.e., a spheroid) and define the
flattening ratio as, $q \equiv c/a$. Thus, the isodensity contours can
be described as:
\begin{equation}
x^{2} + y^{2} + \frac{z^{2}}{q^{2}} = a^{2},
\end{equation}
where $q < 1$ and $q > 1$ are oblate and prolate spheroids,
respectively.  It is widely observed that the inner regions ($R_{gal}
= 5-10 kpc$) of the stellar halo are highly flattened ($q \sim 0.5$),
while the outer regions are more spherically symmetric 
\citep[e.g.,][]{Kinman66,Preston91,Chiba_Beers00}.  We fit
spheroids with various constant flattening to the \loneosI RR Lyrae.
In addition, we also use a model with a spatial variable flattening
ratio \citep{Preston91}, where q varies as function of semi-major
axis, a:
$$ q = \left\{
\begin{array}{cc}
q_{0} + [1 - q_{0}]\frac{a}{20} &\mbox{ if } a < 20 \ kpc \\ 1 &\mbox{
if } a > 20 \ kpc
\end{array}\right.
$$ where $q_{0} = 0.5$ and a is the semi-major axis.  In this variable
q model, the flattening smoothly varies from 0.5 to 1.0 over the
course of 20 kpc. 
Figure  \ref{fig:spherical_and_variable_q_power_laws} shows 
the radial profiles and power law fits
for both the spherical and the variable flattening models.
The power law exponents, \Chisqr, degrees of freedom, and
\Chisqr probabilities are listed in Table \ref{table:flattening}. 
The degrees of freedom are not constant since the data are binned
and the value of $a$ depends on $q$.
All the flattening models have 
larger \Chisqr probability than the spherically
symmetric model indicating that flattened models are favored,  
but there is no clear minimum in the \Chisqr probabilities. 
The variable q model yields a steeper density profile with a power law
exponent of $-3.15 \pm 0.07$ with a \Chisqr per degree of freedom of
0.7.  This agrees with most other RR Lyrae surveys (Table
\ref{table:other_surveys}).
\begin{deluxetable}{lcccccc}
\tablecolumns{7}
\tablewidth{0pc}
\tablecaption{Flattening Fits
\label{table:flattening}}
\tablehead{
\colhead{Sample} & \colhead{Flattening} & \colhead{Exponent} &
\colhead{\Chisqr} & \colhead{Degrees of} &
\colhead{Probability} \\
 & \colhead{Ratio (q)} & & & \colhead{Freedom ($\nu$)} &
\colhead{(Q(\Chisqr,$\nu$))}
}
\startdata
All   &  1.0  &  -2.43 &           24.7  & 24 & 0.42 \\
All   &  Variable q  &  -3.15   &  18.3  & 23 & 0.74 \\  
All   &  0.9  &  -2.48  &          16.1  & 24 & 0.88 \\
All   &  0.8  &  -2.53  &          17.7  & 24 & 0.82 \\ 
All   &  0.7  &  -2.57  &          14.4  & 23 & 0.91 \\ 
All   &  0.6  &  -2.58  &          19.8  & 23 & 0.65 \\ 
All   &  0.5  &  -2.57   &         13.4  & 22 & 0.92 \\
All   &  0.4  &  -2.57   &         15.4  & 22 & 0.84 \\
All   &  1.1  &  -2.35   &         17.5  & 25 & 0.86 \\ 
All   &  1.2  &  -2.29   &         19.1  & 25 & 0.79 \\

\\[-1.4ex] \hline \hline  \\[-1.4ex]

OoI   &  1.0  &  -2.26  &          18.9  &  24 & 0.76 \\ 
OoI   &  Variable q  &  -2.93   &  16.0  &  23 & 0.86 \\ 
OoI   &  0.9  &  -2.29  &          13.8  &  24 & 0.95 \\ 
OoI   &  0.8  &  -2.36  &          14.8  &  24 & 0.93 \\ 
OoI   &  0.7  &  -2.40  &          12.6  &  23 & 0.96 \\ 
OoI   &  0.6  &  -2.42  &          19.8  &  23 & 0.65 \\ 
OoI   &  0.5  &  -2.41   &         10.5  &  22 & 0.98 \\
OoI   &  0.4  &  -2.41   &         10.9  &  22 & 0.98 \\ 
OoI   &  1.1  &  -2.18   &         13.3  &  25 & 0.97 \\ 
OoI   &  1.2  &  -2.15   &         15.9  &  25 & 0.92 \\ 

\\[-1.4ex] \hline \hline  \\[-1.4ex]

OoII  &  1.0  &  -2.88  &          7.2  & 24 & 0.9996 \\ 
OoII  &  Variable q  &  -3.70   &  5.9  & 23 & 0.9998 \\
OoII  &  0.9  &  -2.97  &          5.6  & 24 & 0.9999 \\ 
OoII  &  0.8  &  -3.02  &          5.7  & 24 & 0.9999 \\ 
OoII  &  0.7  &  -3.09  &          7.5  & 23 & 0.9990 \\ 
OoII  &  0.6  &  -3.07  &          5.8  & 23 & 0.9998 \\ 
OoII  &  0.5  &  -3.01   &         9.6  & 22 & 0.9895 \\
OoII  &  0.4  &  -3.04   &         13.0 & 22 & 0.9331  \\
OoII  &  1.1  &  -2.80   &         10.0 & 25 & 0.9966  \\
OoII  &  1.2  &  -2.71   &         11.8 & 25 & 0.9881  \\
 
\enddata
\tablecomments{
Power law fits with varying flattening models. 
}
\end{deluxetable}

The poor constraint on the stellar halo flattening from the \loneosI
RR Lyrae is likely a result of the fact that the Galactic latitude
distribution of these RR Lyrae are
highly concentrated at mid-range Galactic latitude 
with no fields near the Galactic Poles. 
The difference between the spherical and 
flattened model is smallest at lower
Galactic latitudes.

All the recent RR Lyrae surveys are in disagreement with
\citet{Chiba_Beers00} who used $\sim 1200$ non-kinematically selected
metal poor stars. One reason for the difference could be the method
used. RR Lyrae surveys use 3D Galactocentric coordinates to measure
densities. In these surveys, the completeness and survey volume are
relatively well-defined. However, \citet{Chiba_Beers00} use data
derived from many different surveys and is not a volume-limited
sample.  They use a maximum likelihood method which relies on
kinematic information and Jeans' theorem to measure the power
law. Since the completeness and survey volumes are not easily
understood, it is not possible to measure the density distribution
using a binning technique. Thus, it is difficult to understand if
there is a real difference between RR Lyrae stars and generic
metal-poor stars.  In addition, RR Lyrae stars and the metal poor halo
stars from \citet{Chiba_Beers00} may possibly be tracing different
populations. RR Lyrae abundances are influenced by the horizontal
branch morphology of the RR Lyrae progenitor system. Thus, they may
not reveal the entire history of Galactic formation and could have
different spatial distributions than a sample of metal poor halo stars.

\subsection{Period-Amplitude Distribution}
The period-amplitude distribution of RR Lyrae stars can be used to
explore the formation history of the Galactic halo 
\citep[e.g.,][]{Catelan2006}.  If some fraction or all of the stellar halo was
formed from the accretion of dwarf galaxies, which might have very
diverse chemical histories and ages (i.e., different horizontal branch
morphologies), the signatures of these accretion events should be
imprinted in the period-amplitude distribution. Thus, accretion
activity can be probed even after phase-space has been sufficiently
mixed that spatial coherence is lost.

\begin{figure}
\epsscale{1.0}
\centering
\ifsubmode
\plotone{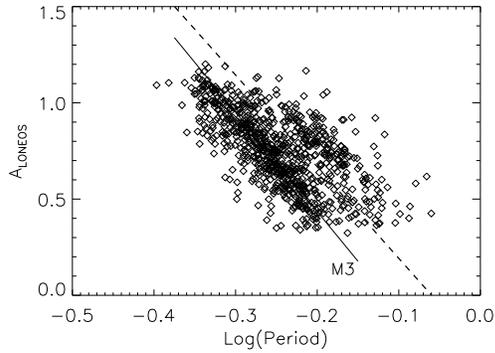}
\fi
\caption[\loneosI period-amplitude distribution with M3 relation]{
The LONEOS RR Lyrae period-amplitude distribution (same as Figure \ref{fig:AFTER_per_amp_cuts}) 
with the linear log(period)-amplitude relation for globular cluster M3 (OoI)
superimposed (solid line).  Note the amplitudes are based on the
unfiltered response of the \loneosI camera system. The M3 relation has
been converted to the \loneosI amplitude scale.  We derived a
conversion between \loneosI amplitude and V-band amplitudes from
\citet{Cacciari2005}.  The dashed line indicates where we divide our
sample into OoI and OoII components.}
\label{fig:loneos_per_amp_M3}
\end{figure}

Our large sample of RR Lyrae stars allows us to further investigate
this possibility and to explore differences in the two components. In
lower-right panel of Figure \ref{fig:AFTER_per_amp_cuts}, 
we plot the period-amplitude
distribution of 838 ab-type RR Lyrae stars.  Since LONEOS uses an
unfiltered CCD system, whose sensitivity peaks in the red, amplitudes
for our RR Lyrae will be systematically smaller than those taken in
the V-band. The observed amplitude of RR Lyrae increases as you
observe them in bluer passbands.\footnote{Note that the transformation
to the $V_{SDSS}$ magnitude system does not affect the measured
\loneosI amplitudes.  The amplitudes are strictly a function of the
passband used.}  Using a subset of \loneosI RR Lyrae (12 stars) which
have previously measured V-band amplitudes taken from SIMBAD database,
we have empirically derived a scaling relation between the LONEOS
amplitudes ($A_{LONEOS}$) and V-band amplitudes ($A^{V}$) and scaled
the M3 RR Lyrae period-amplitude relation to the \loneosI amplitude
scale.  The mean value of $A_{LONEOS}/A^{V}$ is $0.7$ with a standard 
deviation of $0.1$. Figure \ref{fig:loneos_per_amp_M3} 
shows that data from the lower-right panel in  
Figure \ref{fig:AFTER_per_amp_cuts} 
overlaid with the linear period-amplitude
relation for M3 (the quintessential OoI globular cluster)
\citep{Cacciari2005} corrected to the \loneosI amplitude scale. 
Despite only having 12 stars to scale 
LONEOS amplitudes to V-band amplitudes, the
main locus agrees well with the period-amplitude relation of
M3. Thus, we associate the main locus with the OoI component of the
stellar halo. In addition, Figure \ref{fig:loneos_per_amp_M3} shows a
striking concentration of RR Lyrae (26\% of the total sample) with
longer periods offset from the main locus of points.

\begin{figure}
\epsscale{1.0}
\centering
\ifsubmode
\plotone{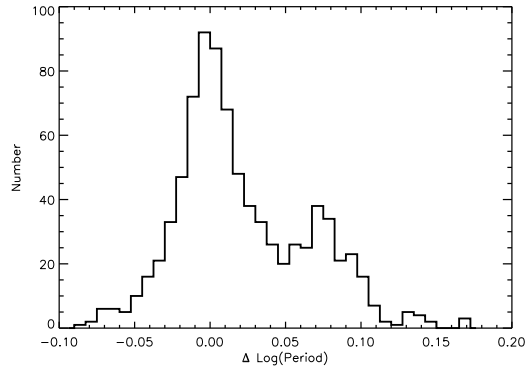}
\fi
\caption[\loneosI Period Shift]{Histogram of the period shift of 
RR Lyrae relative to the primary locus (OoI). We fit a line to the
primary locus and measure $\Delta\ Log(Period)$ which refers to the
Log(Period) distance (at fixed amplitude) of {\em each} RR Lyrae star
from this best fit line.  Thus, the primary locus is centered at
$\Delta\ Log(Period) = 0.0$.  The location of the secondary locus is
located at approximately the location where we would expect the OoII
locus \citep{Sandage1981,Cacciari2005}.  For the subsequent analysis,
we divide our sample at $\Delta\ Log(Period) = 0.045$, where period is
measured in days. }
\label{fig:period_shift}
\end{figure}

We fit a line to the primary locus ($P_{primary}(A)$) and measure the
period shift,
\begin{equation} 
\Delta\ Log(Period) = P_{RR}(A) - P_{primary}(A),
\end{equation}
which refers to the Log(Period) distance (at fixed amplitude) of {\em
each} RR Lyra ($P_{RR}(A)$) from this best fit line ($A = -4.76 \times
Log(P_{primary}) - 0.51$).  Figure \ref{fig:period_shift} shows the
distribution of the period shift at fixed amplitude from the main
locus. This figure clearly shows the existence of a secondary locus.
It is shifted from the main locus by $\Delta Log(P) \sim
0.075$\footnote{Periods are measured in days.}, which is approximately
the shift between RR Lyrae in OoI to OoII globular clusters 
\citep[e.g.,][]{Sandage1981,Clement1999}.  Thus, we associate the
main locus of points with an OoI halo component, and the shifted locus
with an OoII halo component.

\begin{figure}
\epsscale{1.0}
\centering
\ifsubmode
\plotone{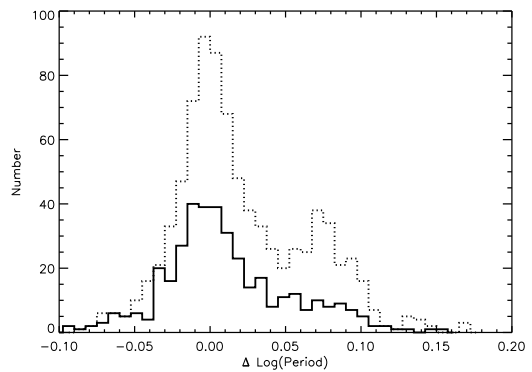}
\fi
\caption[Period shifts of the QUEST-I RR Lyrae Survey]{Period shifts 
(with respect to M3) of the QUEST-I RR Lyrae Survey is shown as a solid
line, while the LONEOS-I period shifts are shown as a dashed line. 
In contrast to the LONEOS-I sample, there is no clearly displaced 
secondary peak associated with an OoII component for the QUEST-I sample. 
However, there is a long period tail.}
\label{fig:quest_period_shift}
\end{figure}

In Figure \ref{fig:quest_period_shift}, we plot the period shift distribution
of 395 RR Lyrae from the QUEST-I catalog \citep{QUEST04}
with mean magnitudes in the range [13.5, 19.7] as well as 
the LONEOS-I period-shift distribution. This QUEST-I distribution 
also contains a sizable fraction ($\sim 20\%$) 
of longer period RR Lyrae than the main OoI locus.
\cite{Catelan2006} has pointed out that the Oosterhoff 
classification the QUEST-I RR Lyrae may be unreliable due to their
sparse temporal sampling. This could cause a smearing of the period shift 
distribution in the QUEST-I data.

\begin{figure}
\epsscale{1.0}
\centering
\ifsubmode
\plotone{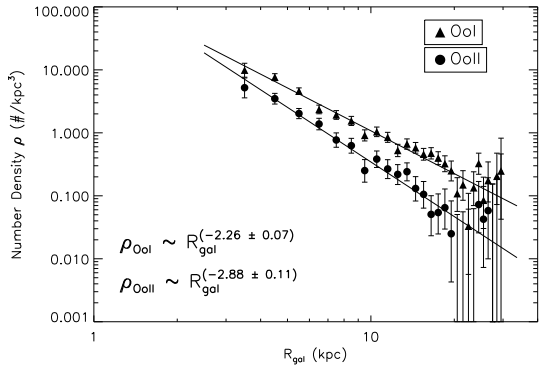}
\fi
\caption[OoI and OoI Radial Distributions]{The spherically symmetric (q=1) 
radial distribution for the OoI and OoII components.  
The bins with zero density are included in the fit; the $1\sigma$ 
Poisson upper limits are calculated using the method of \citet{Gehrels}.
}
\label{fig:radial_above_below}
\end{figure}

It has been well established that OoI and OoII globular clusters show
spatial and kinematic differences, which may signify different origins
\citep{Lee99}. It is natural to ask if the OoI and OoII components of
the stellar halo show similar differences.  We divided our sample into
the two components described above. The division was positioned at the
minimum of the period shift distribution, $\Delta(P) = 0.045$. The
division is denoted by the dashed line in Figure
\ref{fig:loneos_per_amp_M3}. 
The radial number density profiles for each component are shown in
Figure \ref{fig:radial_above_below}. From these
profiles, we derive the following radial dependences:
\begin{equation}
\rho_{OoI} \sim R^{-2.26 \pm 0.07},
\end{equation}

\begin{equation}
\rho_{OoII} \sim R^{-2.88 \pm 0.11}.
\end{equation}
The OoII component has a steeper profile than the OoI component.  This
is consistent with \citet{Lee99} who, using a smaller RR Lyrae sample,
found that the OoI component dominated farther from the Galactic
plane, possibly indicating that the OoII component falls off more
rapidly than the OoI component.  In the following sections, we will
examine whether this difference could be caused by some systematic
effect in the detection of the two populations.

The possibility of an intermediate Oosterhoff halo component still
exists, but our current sample cannot address this.  It is critical to
effectively identify and veto RR Lyrae which exhibit the Blazhko
effect (i.e. amplitude and phase variations). These stars will add
scatter to the period-amplitude relation \citep{Cacciari2005}.  In
addition, we have found that the \loneosI measured RR Lyrae amplitudes
tend to be underestimated (Figure \ref{fig:amp_errors}). This effect
would tend to diminish the number of RR Lyrae with intermediate Oo
classification.  

Additionally, microlensing surveys (e.g., MACHO, OGLE) have found that
as much as a third of RR Lyrae stars in the LMC and SMC exhibit the
Blazhko effect \citep[e.g.,][]{MACHO2000}.  In order to address the
possibilities of an intermediate Oo component, we need better temporal
coverage \citep[e.g.,][]{Catelan2006}. A mixture of OoI and OoII could
masquerade as an intermediate.  

Is the difference in radial distributions associated with the
Oosterhoff effect? Does the Oosterhoff effect manifest itself in such
a way?  Previously, we have chosen to divide our full sample into two
components that resemble OoI and OoII populations; this maximizes the
role of the Oosterhoff effect on the sample. Instead, we now divide
our entire sample along a line in period-amplitude space perpendicular
to Oosterhoff dividing line, minimizing the role of the Oosterhoff
effect.  The resulting power law exponents ($\alpha$) are consistent
with each other ($\alpha = -2.50 \pm 0.07$ and $\alpha = -2.39 \pm
0.1$). This indicates that the Oosterhoff effect is likely associated
with differing radial distributions.

 \section{Discussion}
As described in the previous section, the \loneosI RR Lyrae data have
two interesting characteristics. The Oosterhoff effect is manifest in
the Galactic stellar halo.  Moreover, the two Oosterhoff components
we have identified have distinct radial distributions.

\subsection{Potential Systematic Errors and Biases}
In order to test whether the the Oosterhoff effect in the stellar halo
and difference in the radial distribution of the two Oosterhoff
components is real, we performed a series of tests to examine
systematic effects that might account for the difference. It appears
that this result is robust.

\begin{figure}
\epsscale{1.0}
\centering
\ifsubmode
\plotone{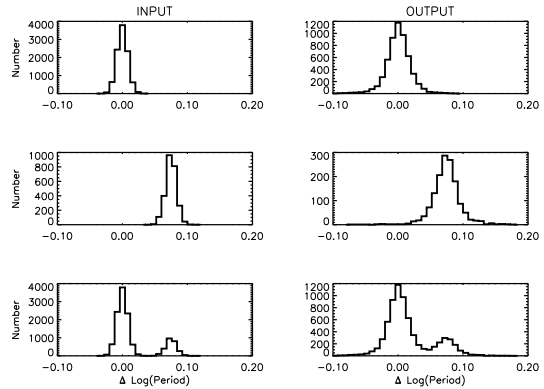}
\fi
\caption[Period shift Histograms of simulated OoI and OoII components]{
We simulated RR Lyrae light curves with period and amplitude distributions of OoI and OoII
RR Lyrae stars. These simulated RR Lyrae light curves were processed through our detection pipeline.
This figure shows the period shift histograms for one field with 28 images (1374B). We have scaled the OoII
input sample to contain $26 \%$ of the OoI component, which is what we observe in the \loneosI
sample. The detection process tends to broaden the period shift distributions and
produces long tails. However, the OoI and OoII are still recognizable
after the detection process.
}
\label{fig:period_shift_OoI_to_OoII}
\end{figure}

\paragraph{Period Errors}
Let us assume that the stellar halo consists of solely an OoI
population of RR Lyrae stars. Is it possible that OoI RR Lyrae stars
might be misidentified as OoII? For example, it is possible that
periods are grossly mis-estimated. Could our timeseries analysis and
detection process artificially produce an OoII population?  We
simulated RR Lyrae light curves with period and amplitude
distributions of OoI and OoII RR Lyrae stars. These simulated RR Lyrae
light curves were processed through our detection pipeline, which was
described in the previous section, and used in the selection of final
\loneosI RR Lyrae sample discussed in this section. Figure
\ref{fig:period_shift_OoI_to_OoII} shows the period shift histograms
for one field with 28 images (1374B). We have scaled the OoII input
sample to contain $26 \%$ of the OoI component, which is what we
observe in the \loneosI sample. The detection process tends to broaden
the period shift distributions and produces long tails. However, the
OoI and OoII components are still recognizable.

\begin{figure}
\epsscale{1.0}
\centering
\ifsubmode
\plotone{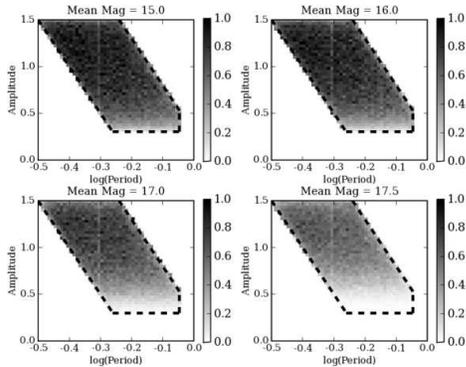}
\fi
\caption[\loneosI Detection Efficiency in Period-Amplitude Space]{
The detection efficiency in period-amplitude space for fields 
with less than 30 images, which are fields with the fewest images.
We simulated 310,000 RR Lyrae stars with mean magnitude of 17.0. 
Note that color table has been inverted (i.e., black corresponds 
to 100\% efficient). We did not find any dependence of the efficieny with 
period at a fixed amplitude.
Also, one can see a deficiency at a period of 0.5 days (-0.3 on the log scale), 
which is caused by the window function of these fields.}
\label{fig:per_amp_complete_Nim_le_29}
\end{figure}

\paragraph{Efficiency}
In order to calculate the RR Lyrae star number density, we need to
estimate the detection efficiencies. This was described in the
previous section.  In calculating the detection efficiencies, we
averaged over the period and amplitude distributions from previous
surveys. These distributions are essentially for the Oosterhoff I
class. The assumption that we have made is that the detection
efficiencies for OoI and OoII are identical.  In order to investigate
differences in the detection efficiencies, we simulate RR Lyrae light
curves to explore the detection efficiencies in the period-amplitude
space. We performed more extensive simulations
examining only fields with less than 30 images, which are fields with
the fewest images and represent the worst-case scenario. We simulated
310,000 RR Lyrae stars with mean magnitudes of 15, 16, 17, \& 17.5
(Figure \ref{fig:per_amp_complete_Nim_le_29}).  There is a strong trend with
amplitude which is expected given the photometric error trend with magnitude.                  
We also observe a deficiency near  a period of 0.5 days (-0.3 on the log scale), 
which is caused by the sampling window function of these fields.  
We find no dependence of the efficieny with 
period at fixed amplitude. Thus, we are confident that 
the efficiency differences do not account for the difference in 
the the radial distributions of the OoI and OoII components.

What effect could an error in the detection efficiency have? We have
measured the power law indices for two Oosterhoff components without
including the detection efficiency correction. This yields the
following fits:
\begin{equation}
\rho_{OoI} \sim R^{-2.68 \pm 0.06},
\end{equation}

\begin{equation}
\rho_{OoII} \sim R^{-3.17 \pm 0.1}.
\end{equation}
The detection efficiency correction tends to make the distributions
shallower. Even if we take the extreme scenario where our detection
efficiency correction is only applied to the OoII component and we
assume that the OoI component has a 100\% detection efficiency, the
difference in the radial distributions of the Oosterhoff components
remains statistically significant.

\begin{figure}
\epsscale{1.0}
\centering
\ifsubmode
\plottwo{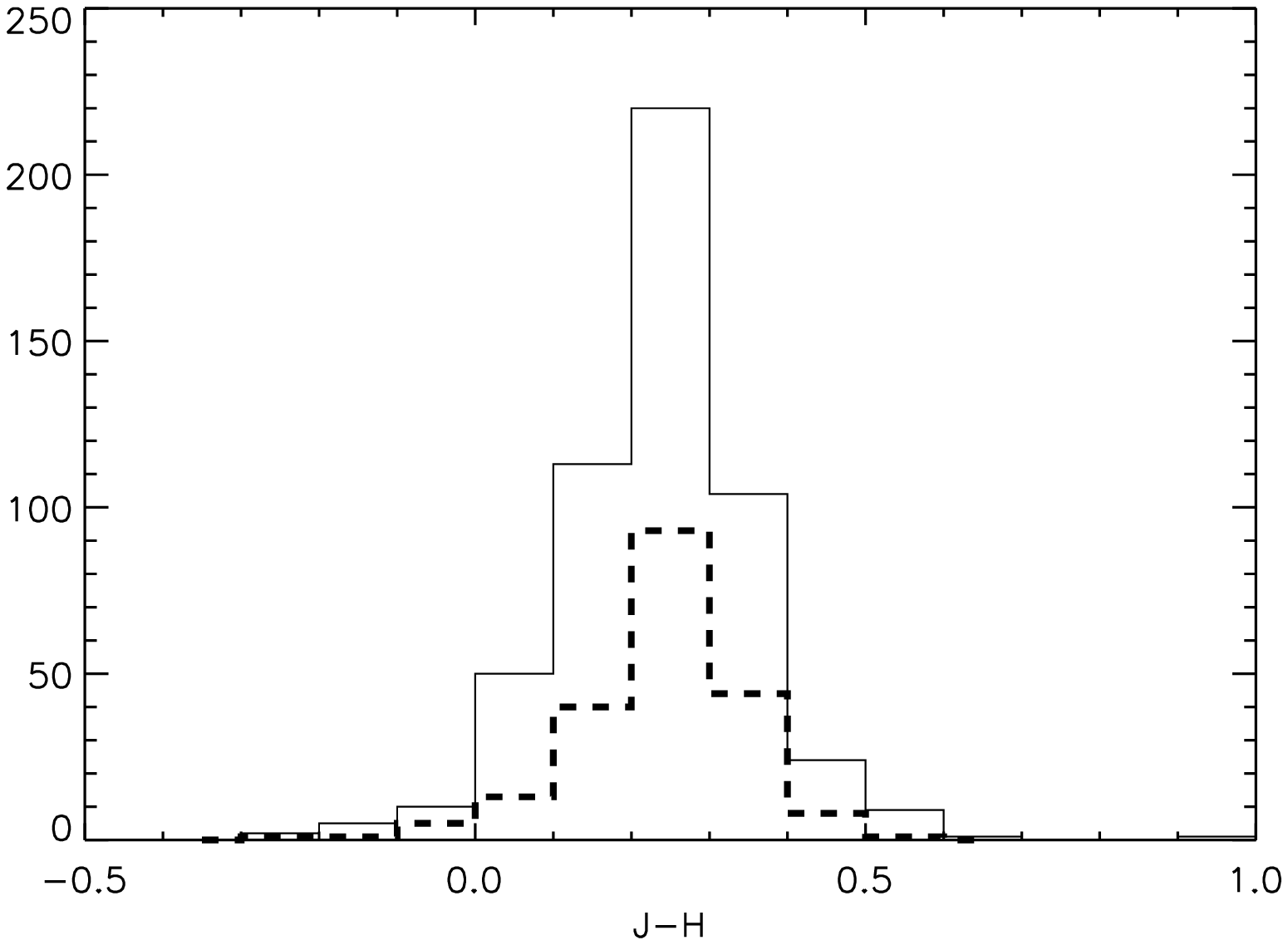}{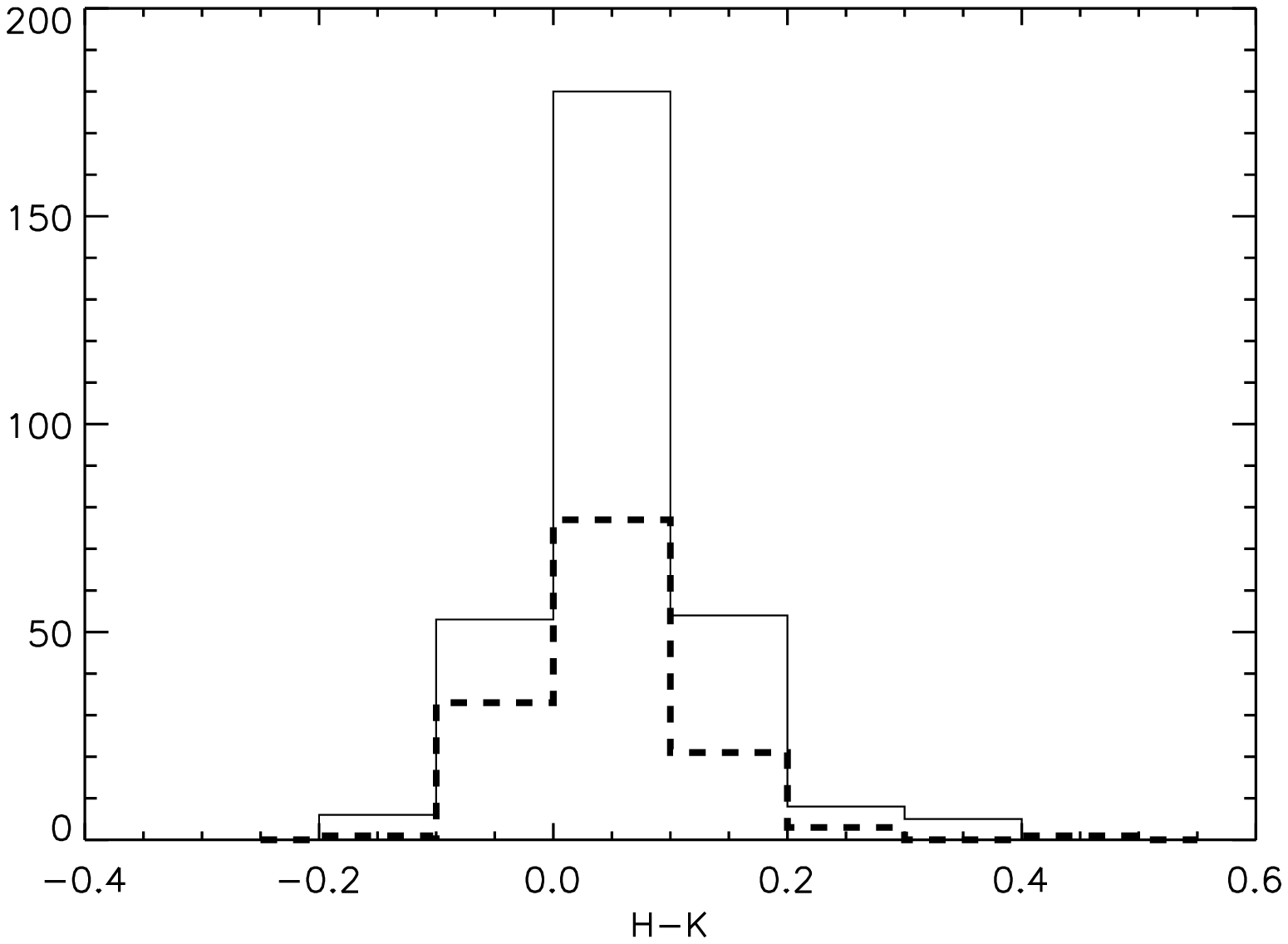}
\fi
\caption[2MASS colors for the OoI and OoII components.]{
2MASS colors for the OoI and OoII components.  There is no 
significant color difference between the two components. }
\label{fig:above_below_2mass_colors}
\end{figure}

\begin{figure}
\epsscale{1.0}
\centering
\ifsubmode
\plotone{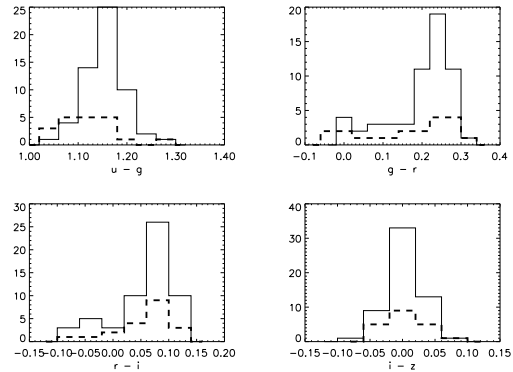}
\fi
\caption[SDSS colors for the OoI and OoII components.]{SDSS color distributions 
for \loneosI RR Lyrae stars (78). The Oo I component is shown as the solid curve,
while the  Oo II component is shown as a dashed line.}
\label{fig:above_below_sdss_colors}
\end{figure}

\paragraph{Color Differences}
Is it possible to explain the OoI and OoII difference because of color
difference? Does the photometric calibration introduce a bias based on
color? We looked to see if our OoI and OoII components have different
color distributions. First, we have acquired dereddened 2MASS colors
for several hundred of the \loneosI RR Lyrae stars. 
Figure \ref{fig:above_below_2mass_colors} show the color distributions for
each Oosterhoff component. The distributions appear to be consistent
and no infrared color difference is seen. In addition, Figure
\ref{fig:above_below_sdss_colors} shows the color distributions for
the 78 RR Lyrae stars with SDSS DR3 photometry.  While this sample is
small, we do not see major differences in the SDSS colors of the
\loneosI RR Lyrae stars, except for the $u - g$ color, where the OoI
component appears to be bluer. This  $u - g$ color difference could be due
to a difference in metallicity.

\begin{figure}
\epsscale{1.0}
\centering
\ifsubmode
\plotone{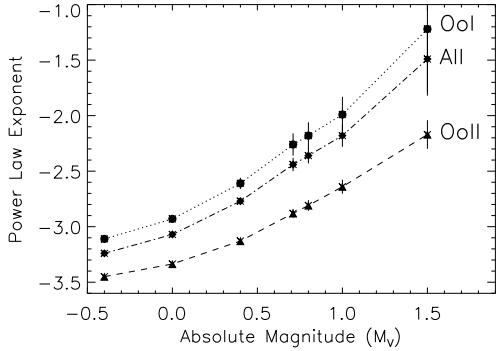}
\fi
\caption[Power Law Exponents as a function of RR Lyrae Absolute Magnitude]{
The dependence of the power law exponent on the RR Lyrae absolute magnitude for
the entire LONEOS-I RR Lyrae sample, OoI and OoII components is shown. 
If we fix the absolute magnitude of the OoI
stars to $M_{V} = 0.71$, this would require that the absolute
magnitude of the OoII component be roughly $M_{V} > 1.5 $. 
This value is not within the observed range of $M_{V}$ for RR Lyrae of any kind.
Thus, an absolute magnitude difference between the
OoI and OoII components cannot explain the different radial
distributions.
}
\label{fig:absolute_mag}
\end{figure}

\paragraph{Zeropoint Differences}
Zeropoint differences between the two Oosterhoff components may arise
in several forms: absolute magnitude differences, photometric
zeropoint offsets, and mean magnitudes errors. First, we explore
whether the difference in the radial distribution of the OoI and the
OoII components could result from intrinsically different absolute
magnitudes.  In Figure \ref{fig:absolute_mag}, we show the dependence
of the power law exponent on the absolute magnitude for the two
Oosterhoff components. If we fix the absolute magnitude of the OoI
stars to $M_{V} = 0.71$, this would require that the absolute
magnitude of the OoII component be roughly $M_{V} \ge 1.5$ 
in order to have the same power law slope. This value is 
not within the observed range of $M_{V}$ for RR Lyrae of any kind.
Thus, an absolute magnitude difference between the
OoI and OoII components cannot explain the different radial
distributions.

Does the photometric calibration introduce a bias based on color?
Even though we do not see a clear color difference between the
Oosterhoff components, let us explore how a difference could affect
our results. As we saw in Section 2 (Figure
\ref{fig:loneos_sdss_color_dependence}), the offset from \loneosI
magnitudes (tied to GSC2 F) to $V_{SDSS}$ is a function of color.  A
systematic error in this offset directly translates to a systematic
offset in the mean magnitudes (and distances) of all the RR Lyrae
stars. This will lead to different radial profiles.
We have used an offset for a color range of RR Lyrae stars. Figure
\ref{fig:loneos_sdss_color_dependence} indicates that the offset may
vary by $\sim 0.3$ at the extremes of the RR Lyrae color range. To
give us an idea of how this affects the radial profiles of the
Oosterhoff components, let us look at Figure \ref{fig:absolute_mag},
where we explore how changes in the absolute magnitude affect the
radial distributions. An overall shift in the \loneosI zeropoint is
essentially changing the RR Lyrae absolute magnitude.  If we fix
$M_{V}$ and assume the two Oosterhoff components have a $0.2$ mag
overall zeropoint difference, this can not explain the difference in
the two radial distributions.  One needs an overall zeropoint
difference of $\sim 1$ mags, which we are confident does not exist.

It is possible that intrinsic light curve shape differences in the OoI
and OoII components could result in a systematic bias in the mean
magnitudes of the OoI and OoII components.  Any systematic difference
in the measured mean magnitude is a zeropoint offset.  However, Figure
\ref{fig:absolute_mag} shows that the mean magnitude bias would have
to be a sizable fraction of the RR Lyrae amplitude and thus
unrealistic that the radial profiles could result from a bias in the
measure of the mean magnitudes.

\begin{figure}
\epsscale{1.0}
\centering
\ifsubmode
\plotone{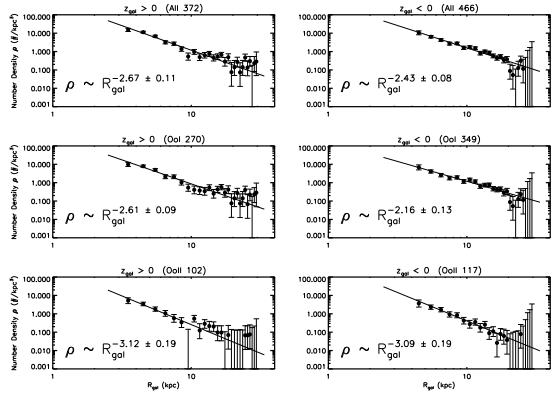}
\fi
\caption[Radial density profiles above and below the Galactic Plane]{
We have divided the \loneosI RR Lyrae sample into regions above and below the Galactic plane.
This figure shows the spherically symmetric radial distributions
for the entire sample as well as for the OoI and OoII components. The OoII component's profiles
are consistent with the full OoII sample. However, the OoI shows a 2.8 $\sigma$ difference above
and below the Galactic plane with power law exponents of $-2.61 \pm 0.09$ and $-2.16 \pm 0.13$.
If this difference is real, it could indicate spatial asymmetries
in the OoI component. However, the above and below plane difference in
power law slope is of marginal statistical significance.
}
\label{fig:above_below_plane_power_laws}
\end{figure}

\paragraph{Spatial Sub-Samples}
We have divided the \loneosI RR Lyrae sample into regions above and
below the Galactic plane.  This division roughly divides the entire
sample in half (372 and 466, above and below the plane, respectively).
Figure \ref{fig:above_below_plane_power_laws} shows the spherically
symmetric radial distributions for the entire sample as well as for
the OoI and OoII components (see also Table~\ref{table:above_below_disk}). The OoII component's profiles are
consistent with the full OoII sample. However, the OoI shows
a 2.8 $\sigma$ difference above and below the Galactic plane with power 
law exponents of $-2.61 \pm 0.09$ and $-2.16 \pm 0.13$, respectively. 
If this difference is real, it could indicate spatial asymmetries 
in the OoI component. However, the above and below plane difference in 
power law slope is of marginal statistical significance.
We are engaged in the analysis of the LONEOS II data set 
which (with its much greater sky coverage) should lay this issue 
to rest one way or another.
It is also possible that this difference is caused by some systematic variation in
the \loneosI photometric completeness function which we have not taken
into account. Recall, that we use a global photometric completeness
function for \loneosI and a field-by-field temporal completeness
correction. We have investigated if there was a systematic difference between  
the \loneosI fields above and below the Galactic plane.
First, we derived a photometric completeness function for the \loneosI fields above 
and below the plane using SDSS Data Release 5. Using the same criteria that we used 
to obtain the  global photometric completeness function 
(Figure  \ref{fig:dr1_phot_complete}), we queried the \loneosI database 
for a detection in each image in the database and computed photometric 
completeness functions above and below the Galactic Plane. There were about 
one million detections of SDSS DR5 stars in the \loneosI database 
above and below the Galactic Plane. However, we did not find a significant  
difference between photometric completeness functions 
above and below the Galactic Plane. Next, we performed lightcurve 
simulations, like those in Figure \ref{fig:per_amp_complete_Nim_le_29}, 
to compare the temporal detection efficiency in the period-amplitude space for 
\loneosI fields above and below the Galactic plane.  Again, no significant
difference was seen for the fields above and below the Galactic plane.
However, the OoI and OoII still have different profiles in each
hemisphere, but the extent of the difference is not presently clear.
\begin{deluxetable}{lcc}
\tablecolumns{3}
\tablewidth{0pc}
\tablecaption{Power law fits above and below the Galactic Disk
\label{table:above_below_disk}}
\tablehead{
\colhead{Location} & \colhead{Sample (Size)} &  \colhead{Exponent} \\ }
\startdata
$z_{gal} > 0$  &  All (372)  &   $ -2.67 \pm 0.11 $       \\
$z_{gal} < 0$  &  All (466)  &   $ -2.43 \pm 0.08 $       \\
All            &  All (838)  &   $ -2.43 \pm 0.06  $       \\

$z_{gal} > 0$  &  OoI (270)  &   $ -2.61 \pm 0.09 $       \\
$z_{gal} < 0$  &  OoI (349)  &   $ -2.16 \pm 0.13 $       \\
All            &  OoI (619)  &   $ -2.26 \pm 0.07  $       \\

$z_{gal} > 0$  &  OoII (102)  &  $ -3.12 \pm 0.19 $       \\
$z_{gal} < 0$  &  OoII (117) &   $ -3.09 \pm 0.19 $       \\
All            &  OoII (219) &   $ -2.88 \pm 0.11$        \\ 

\enddata
\end{deluxetable}

\paragraph{CCD Focal Plane Distribution}
In order to obtain the wide field of view needed for LONEOS, the
telescope used has a Schmidt design. A Schmidt telescope consists of a
spherical primary mirror and a corrector lens at its radius of
curvature.  However, distortions in the focal plane persist. These
distortions increase near the edges of the focal plane (i.e., with
increased distance from the optical axis).  Such distortion might affect
the amplitude of our RR Lyrae stars. We examined the distribution of
our RR Lyrae stars in the CCD focal plane to search for any unusual
clustering that might result from the optical distortions. We use an 
extension of a one-dimensional Kolmogorov-Smirnov test to two-dimensions \citep{NR}.
We find that the RR Lyrae distribution is not 
significantly different from a random 2D distribution.  
In addition, we examine the distributions of the two Oosterhoff components.  
The KS probability indicates that these distributions are not 
significantly different from one another.

\paragraph{Large Amplitude Sample}
The detection efficiency generally decreases with decreasing
amplitude.  We examined the radial distributions of the OoI and OoII
components while restricting the sample to include only RR Lyrae stars
with amplitudes greater than 0.6. The difference in the radial
distributions of the two components remains unchanged.

\paragraph{Increased Cadence Subsample}
Finally, we look at a subsample selected from fields with more than 32
epochs.  Such a subsample should be less affected by efficiency
differences.  This results in 370 OoI and 129 OoII RR Lyrae stars. The
power law exponents are $\alpha = -2.46 \pm 0.1$ and $\alpha = -2.95
\pm 0.14$ for the OoI and OoII components, respectively. Again, the
difference in radial distributions remains.

\subsection{Implications for Galactic Formation}
The first serious model of galaxy formation was proposed by
\citet{ELS} (ELS). Based on correlations between metallicity,
orbital eccentricity, angular momentum, and
velocity perpendicular to the Galactic Plane  in a sample
of nearby, high velocity (halo) field stars, they concluded
that the Galaxy formed from the rapid (within the free fall time), 
monolithic collapse of an
overdensity in the primordial density field.
The first stars that formed during the collapse
inherited the low metallicity and predominantly radial motion
of the collapsing gas, while conservation of angular momentum and
energy dissipation led to the formation of a flattened and circularly
rotating thin disk of gas.
Metal-rich stars would form in this disk as
the interstellar medium became enriched with the products of
nucleosynthesis during stellar evolution.

There has been mounting evidence to suggest that the ELS scenario
does not agree with observations. \citet{SZ} (SZ)  presented an extensive
survey of globular clusters in the outer halo that
showed a spread in their horizontal branch (HB) morphology
(i.e., color distribution of stars along the HB) that is
independent of metallicity, the ``second parameter problem''.
This effect was not evident in clusters with smaller galactocentric
distances.  SZ proposed that age was the second parameter and that the
outer clusters originated in satellite
systems that were subsequently accreted by the Milky Way.
The spread in HB morphology can be explained by different
star formation histories in the accreted systems.
Today, the most successful large-scale 
structure formation models are  based on hierarchical clustering (i.e., the 
Cold Dark Matter (CDM) paradigm).  The SZ scenario for galaxy 
formation is attractive since it would fit within the CDM paradigm
of structure formation.  
Several accreted system have been detected in the Milky Way 
\citep[e.g.,][]{Ibata01, Belokurov2007}, which would provide 
direct support for SZ scenario.  
 
In studies of globular cluster RR Lyrae stars, the location of an RR
Lyra star in period-amplitude space has been correlated to metallicity
\citep[e.g.,][]{Sandage93,Contreras2005} and age relative to the
Zero-Age Horizontal Branch (ZAHB)
\citep[e.g.,][]{Lee1990,Clement1999}.  \citet{Clement1999},
\citet{Lee99}, and \citet{Jurcsik2003} conclude that the Oosterhoff
effect is caused by evolution off the ZAHB. They found that OoII
globular clusters are not correlated with metallicity, but rather
their RR Lyrae stars appear to have evolved off the ZAHB.  Relative
age estimates between the OoI and OoII globular cluster populations
are 2-3 Gyr \citep{Lee99}.  \citet{Lee99} have found that the
kinematic properties of OoI and OoII globular clusters are
different. They found that the OoI clusters have little or no net
angular momentum, which points to a formation scenario like that of SZ. 
On the other hand, OoII clusters were found to have a net
prograde rotation ($V_{rotation} = -94 \pm 47$ km/s), which indicates
that their formation scenario resembles the ELS scenario.
The kinematic data and the relative age differences (indicated by the
Oosterhoff type) of OoI and OoII clusters point to a dual mode
formation scenario for the globular clusters. 

Our detection of the Oosterhoff effect in the Galactic stellar 
halo indicates that the formation of the globular clusters 
may be connected to that of the Galactic stellar halo.  
If the Oosterhoff type indicates an age
difference, then the Oosterhoff effect in the stellar halo suggests a
time gap between the formation of the OoI and OoII components.  It
should be noted that an age difference between OoI and OoII components
does not give us any information about when the OoI component may have
been accreted into the Milky Way. However, the presence of the two
distinct components in the stellar halo discovered in \loneosI
indicates that the progenitors of the stellar halo had diverse ages
and/or chemical properties.

One way to connect the OoI and OoII components 
to the ELS and SZ scenarios would
be to search for phase-space substructure. First, if the OoI component
was formed predominantly in the SZ scenario than we would expect the
stellar halo to have over- and under-densities.  While there has been
much excitement concerning the discovery of the overdensities in the
stellar halo that are associated with accreted systems, the stellar
halo should also have voids or underdensities if it was formed in an
SZ scenario.  The radial distributions of both components exhibit no
significant substructure.  However spatial substructure tends to be
quickly erased. Thus, the lack of substructure in the OoI component
may simply indicate that the number of accretions in the past 1-2 Gyrs
has been low. In addition, \loneosI probes a small fraction of the 
Galactic Halo. A much larger survey is needed to explore whether there are 
differences in the amount of substructure in the OoI and OoII components.

Besides the possible time gap between the formation
of the OoI and OoII components, the difference in the radial
distributions of the OoI and the OoII components is suggestive of
multiple physical mechanisms for the formation of the stellar halo.
If one assumes that the Galactic stellar halo was formed by a
combination of ELS and SZ scenarios, then one might associate the two
Oosterhoff components of the stellar halo to each of those scenarios.
The relatively steeper radial distribution of the OoII component as
compared to the OoI component could be a manifestation of the fact
that an ELS scenario is more dissipative than an SZ scenario.  An ELS
scenario would be more dissipative since there is much gas in the
system. Loss of energy (e.g.,  from radiative transfer phenomena) would
result in a contraction of the halo. Whether the contraction was
isotropic or not depends on the amount of angular momentum in the
system.  


However, our detection of the Oosterhoff effect in the stellar halo
does not necessarily imply that the stellar halo was 
formed through a combination of ELS and SZ scenarios.
It might be able to form two radially
distinct population of RR Lyrae stars through a combination 
of hierarchical mergers which could have been dissipative and
dissipationless \citep[e.g.,][]{Bekki_chiba2001,Chiba_Beers00}. 
It is possible that systems accreted at early times 
contained considerable gas, and
would have resulted in dissipative merging. However, systems
accreted at later times would have contained less gas due to star formation 
and ejection from supernovae explosions. This late-time merging 
would have resulted in dissipationless merging. 
Using a large sample of stellar spectra, \citet{Carollo_Beers07} 
found that the stellar halo is composed of at least two spatially, 
kinematically, and chemically distinct components, 
which agrees with our results. \citet{Carollo_Beers07} 
argue that their data can be explained solely within the 
context of hierarchical merging. This scenario could account
for the two spatially distinct components in the \loneosI sample as well.

We have found that the stellar halo is predominately OoI, 
with a significant OoII component. We cannot rule out whether 
the stellar halo contains an
intermediate Oosterhoff class. 
\citet{Siegel2006} and \citet{Bootes_DallOra_2006} have found
that the recently discovered Bo{\"o}tes dSph galaxy to be of OoII type.  
Since the Bo{\"o}tes dSph is metal-poor and is an OoII, it 
supports the argument that the Oosterhoff effect is a continuous 
effect which is predominately associated with metallicity.  
This new finding raises the possibility the OoII component of the Galactic
stellar halo was formed from accreted systems like Bo{\"o}tes. 

The chemical abundance properties of the stellar halo and the
dwarf galaxies are distinct \citep[e.g.,][]{Venn2004}. The stellar
halo is more metal-poor, but with enhanced $[\alpha/Fe]$ abundances,
than the Local Group dwarf galaxies.  \citet{Robertson2005} have shown
that such a discrepancy can be explained within the CDM 
paradigm. Using cosmologically-motivated star formation and gas
accretion histories, they argue that most of the stellar halo was
formed from short-lived satellites which were accreted and destroyed
at early times.  
Since these early progenitor satellites were
short-lived, there is limited time for stellar enrichment and the
dominant contributor to the enrichment is from Type II supernovae
(i.e., $\alpha$-enrichment).  However, the present day Local Group
dwarf galaxies have had a long time to produce metals from both Type I
and II supernovae. The bottom line is that the dwarf galaxies that we
observe in the Local Group may not be representative of the
progenitors of the stellar halo.  A detailed study of metallicity of
the two Oosterhoff halo components will allow us determine if there is 
an age difference between these two Oo components.

\section{Conclusions}

We present a catalog of 838 {\it ab}-type RR Lyrae in the Galactic
stellar halo from the \loneosI survey. We find evidence for two
distinct components in this sample. We associate these two components
with a manifestation of the Oosterhoff effect in the stellar halo. In
addition, we find that these two components have significantly
different radial distributions. Without flattening, the OoI has a
power-law exponent of $-2.26 \pm 0.1$ while the OoII component has a
significantly steeper slope with $-2.88 \pm 0.04$. The flattening of
the halo cannot be constrained by this data set since the distribution
on the sky is limited. We performed numerous tests and simulations to
verify the existence of the two components and their difference in
their Galactocentric radial distributions. Our finding is robust with
respect to potential systematic errors due to e.g. detection
efficiencies, zeropoint problems, color differences, camera
systematics, etc.  It is likely that the OoII component potentially 
represents an older population and traces an early dissipative collapse 
or early dissipative merging activity. 
The OoI component represents the later accretion of younger
systems. Spectroscopic data would be very valuable, the
radial velocities will tell us if these OoI and OoII components have
different kinematic properties and occupy different regions in
phase-space. Spectroscopic data could also supply evidence for
chemical differences. Chemical differences in these two populations
will give us information regarding the environment and age of these
populations. 

In 2000, the LONEOS camera (\loneosII) was upgraded, 
and now consists of two 2k x 4k backside-illuminated CCDs, 
giving it an 8.1 square degree FOV (2.8 arcsec/pixel), 
and reaches to a depth of $R \sim 19.3$.
Approximately one quarter of the sky is covered with more than 40 epochs of
measurements.   Photometric reduction and analysis of the 
new camera data has commenced, and
results will be discussed in future papers. 
The \loneosII data should also increase our RR Lyrae
sample by an order of magnitude with a wider and deeper survey
area. The data from the \loneosII camera
will produce cleaner period-amplitude distributions.  This will be
enable us to search for additional ``fine structure'' in the
period-amplitude distribution.

\section{Acknowledgments}
AR is grateful to the NOAO Goldberg fellowship program for its support.
CWS is grateful for the support from the Packard Foundation that was 
essential to the success of this project. CWS and AM are also 
grateful to Harvard University for their support. 
KHC's work was performed under the auspices of the U.S.
Department of Energy, National Nuclear Security Administration by the
University of California, Lawrence Livermore National Laboratory under
contract No. W-7405-Eng-48.
Conversations with Scott Anderson, Al Diercks, Steven Majewski, 
Gajus Miknaitis, Nick Suntzeff, George Wallerstein, 
and Doug Welch provided encouragement, insight and motivation 
for this work. We also acknowledge the anonymous referee's 
comments which made this a better paper.
The LONEOS project is support by the Lowell Observatory and NASA 
grant NAGW-3397.
This research has made use of the SIMBAD database, operated at CDS, 
Strasbourg, France.
Simulations in this paper were performed with the use of Condor, 
high-throughput distributed computing system at the University of Washington. 
The Condor  Software Program (Condor) was developed by the Condor Team at the
Computer Sciences Department of the University of Wisconsin-Madison.
All rights, title, and interest in Condor are owned by the Condor Team
(\url{http://www.cs.wisc.edu/condor/}).


\clearpage
\bibliographystyle{apj}
\bibliography{ms}

\begin{thebibliography}{}

\bibitem[\protect\citeauthoryear{{Abazajian} et~al.}{{Abazajian}
  et~al.}{2005}]{SDSS_DR3}
{Abazajian}, K.,  et~al. 2005, \aj, 129, 1755

\bibitem[\protect\citeauthoryear{{Alcock} et~al.}{{Alcock}
  et~al.}{2000}]{MACHO2000}
{Alcock}, C., et~al. 2000, \apj, 542, 257

\bibitem[\protect\citeauthoryear{{Alcock} et~al.}{{Alcock}
  et~al.}{1998}]{MACHO_Bulge_RR}
{Alcock}, C., et~al. 1998, \apj, 492, 190

\bibitem[\protect\citeauthoryear{{Alcock} et~al.}{{Alcock}
  et~al.}{1995}]{MACHO1995}
{Alcock}, C., et~al. 1995, \aj, 109, 1653

\bibitem[\protect\citeauthoryear{{Bekki} \& {Chiba}}{{Bekki} \&
  {Chiba}}{2001}]{Bekki_chiba2001}
{Bekki}, K.,  \& {Chiba}, M. 2001, \apj, 558, 666

\bibitem[\protect\citeauthoryear{{Belokurov} et~al.}{{Belokurov}
  et~al.}{2007}]{Belokurov2007}
{Belokurov}, V.,  et~al. 2007, \apj, 654, 897

\bibitem[\protect\citeauthoryear{{Benedict} et~al.}{{Benedict}
  et~al.}{2002}]{Benedict02}
{Benedict}, G.~F.,  et~al. 2002, \aj, 123, 473

\bibitem[\protect\citeauthoryear{{Bowell} et~al.}{{Bowell}
  et~al.}{1995}]{Bowell1995}
{Bowell}, E., {Koehn}, B.~W., {Howell}, S.~B., {Hoffman}, M.,  \& {Muinonen},
  K. 1995, Bulletin of the American Astronomical Society, 27, 1057

\bibitem[\protect\citeauthoryear{{Cacciari}, {Corwin}, \& {Carney}}{{Cacciari}
  et~al.}{2005}]{Cacciari2005}
{Cacciari}, C., {Corwin}, T.~M.,  \& {Carney}, B.~W. 2005, \aj, 129, 267

\bibitem[\protect\citeauthoryear{{Carney} et~al.}{{Carney}
  et~al.}{1995}]{Carney_Galactic_Center}
{Carney}, B.~W., {Fulbright}, J.~P., {Terndrup}, D.~M., {Suntzeff}, N.~B.,  \&
  {Walker}, A.~R. 1995, \aj, 110, 1674

\bibitem[\protect\citeauthoryear{{Carollo} et~al.}{{Carollo}
  et~al.}{2007}]{Carollo_Beers07}
{Carollo}, D., et~al. 2007, astro-ph/07063005

\bibitem[\protect\citeauthoryear{{Catelan}}{{Catelan}}{2006}]{Catelan2006}
{Catelan}, M. 2006, astro-ph/0604035

\bibitem[\protect\citeauthoryear{{Chiba} \& {Beers}}{{Chiba} \&
  {Beers}}{2000}]{Chiba_Beers00}
{Chiba}, M.,  \& {Beers}, T.~C. 2000, \aj, 119, 2843

\bibitem[\protect\citeauthoryear{{Clement} et~al.}{{Clement}
  et~al.}{2001}]{Clement2001}
{Clement}, C.~M., et~al. 2001, \aj, 122, 2587

\bibitem[\protect\citeauthoryear{{Clement} \& {Shelton}}{{Clement} \&
  {Shelton}}{1999}]{Clement1999}
{Clement}, C.~M.,  \& {Shelton}, I. 1999, \apjl, 515, L85

\bibitem[\protect\citeauthoryear{Contreras et~al.}{Contreras
  et~al.}{2005}]{Contreras2005}
Contreras, R., Catelan, M., Smith, H.~A., Pritzl, B.~J.,  \& Borissova, J.
  2005, Astrophys. J., 623, L117

\bibitem[\protect\citeauthoryear{{Dall'Ora} et~al.}{{Dall'Ora}
  et~al.}{2006}]{Bootes_DallOra_2006}
{Dall'Ora}, M., et~al. 2006, \apjl, 653, L109

\bibitem[\protect\citeauthoryear{{Diercks} et~al.}{{Diercks}
  et~al.}{1995}]{Diercks1995}
{Diercks}, A.~H., {Angione}, J., {Stubbs}, C.~W., {Cook}, K.~H., {Bowell}, E.,
  {Koehn}, B.~W., {Nye}, R.~A.,  \& {Dodgen}, D. 1995, in Proc. SPIE Vol. 2416,
  p. 58-64, Cameras and Systems for Electronic Photography and Scientific
  Imaging, Constantine N. Anagnostopoulos; Michael P. Lesser; Eds., 58

\bibitem[\protect\citeauthoryear{{Eggen}, {Lynden-Bell}, \& {Sandage}}{{Eggen}
  et~al.}{1962}]{ELS}
{Eggen}, O.~J., {Lynden-Bell}, D.,  \& {Sandage}, A.~R. 1962, \apj, 136, 748

\bibitem[\protect\citeauthoryear{{Gehrels}}{{Gehrels}}{1986}]{Gehrels}
{Gehrels}, N. 1986, \apj, 303, 336

\bibitem[\protect\citeauthoryear{{Hawkins}}{{Hawkins}}{1984}]{Hawkins84}
{Hawkins}, M.~R.~S. 1984, \mnras, 206, 433

\bibitem[\protect\citeauthoryear{{Heck}}{{Heck}}{1988}]{Heck88}
{Heck}, A. 1988, \aaps, 75, 237

\bibitem[\protect\citeauthoryear{{Ibata} et~al.}{{Ibata}
  et~al.}{2001}]{Ibata01}
{Ibata}, R., {Lewis}, G.~F., {Irwin}, M., {Totten}, E.,  \& {Quinn}, T. 2001,
  \apj, 551, 294

\bibitem[\protect\citeauthoryear{{Ivezi{\' c}} et~al.}{{Ivezi{\' c}}
  et~al.}{2000}]{Ivezic00}
{Ivezi{\' c}}, {\v Z}.,  et~al. 2000, \aj, 120, 963

\bibitem[\protect\citeauthoryear{{Ivezi{\' c}} et~al.}{{Ivezi{\' c}}
  et~al.}{2005}]{Ivezic2005}
{Ivezi{\' c}}, {\v Z}., {Vivas}, A.~K., {Lupton}, R.~H.,  \& {Zinn}, R. 2005,
  \aj, 129, 1096

\bibitem[\protect\citeauthoryear{{Jurcsik} et~al.}{{Jurcsik}
  et~al.}{2003}]{Jurcsik2003}
{Jurcsik}, J., {Benk{\H o}}, J.~M., {Bakos}, G.~{\' A}., {Szeidl}, B.,  \&
  {Szab{\' o}}, R. 2003, \apjl, 597, L49

\bibitem[\protect\citeauthoryear{{Kinemuchi} et~al.}{{Kinemuchi}
  et~al.}{2006}]{Kinemuchi2006}
{Kinemuchi}, K., {Smith}, H.~A., {Wo{\'z}niak}, P.~R.,  \& {McKay}, T.~A. 2006,
  \aj, 132, 1202

\bibitem[\protect\citeauthoryear{{Kinman}, {Wirtanen}, \& {Janes}}{{Kinman}
  et~al.}{1966}]{Kinman66}
{Kinman}, T.~D., {Wirtanen}, C.~A.,  \& {Janes}, K.~A. 1966, \apjs, 13, 379

\bibitem[\protect\citeauthoryear{{Krisciunas}}{{Krisciunas}}{2001}]{KK_thesis}
{Krisciunas}, K. 2001, Ph.D.~Thesis (University of Washington)

\bibitem[\protect\citeauthoryear{{Krisciunas}, {Margon}, \&
  {Szkody}}{{Krisciunas} et~al.}{1998}]{KK98}
{Krisciunas}, K., {Margon}, B.,  \& {Szkody}, P. 1998, \pasp, 110, 1342

\bibitem[\protect\citeauthoryear{{Layden}}{{Layden}}{1995}]{Layden1995}
{Layden}, A.~C. 1995, \aj, 110, 2288

\bibitem[\protect\citeauthoryear{{Layden}}{{Layden}}{1998}]{Layden1998}
{Layden}, A.~C. 1998, \aj, 115, 193

\bibitem[\protect\citeauthoryear{{Layden} et~al.}{{Layden}
  et~al.}{1996}]{Layden96}
{Layden}, A.~C., {Hanson}, R.~B., {Hawley}, S.~L., {Klemola}, A.~R.,  \&
  {Hanley}, C.~J. 1996, \aj, 112, 2110

\bibitem[\protect\citeauthoryear{{Lee} \& {Carney}}{{Lee} \&
  {Carney}}{1999}]{Lee99}
{Lee}, J.,  \& {Carney}, B.~W. 1999, \aj, 118, 1373

\bibitem[\protect\citeauthoryear{{Lee}, {Demarque}, \& {Zinn}}{{Lee}
  et~al.}{1990}]{Lee1990}
{Lee}, Y., {Demarque}, P.,  \& {Zinn}, R. 1990, \apj, 350, 155

\bibitem[\protect\citeauthoryear{{Oort} \& {Plaut}}{{Oort} \&
  {Plaut}}{1975}]{Oort75}
{Oort}, J.~H.,  \& {Plaut}, L. 1975, \aap, 41, 71

\bibitem[\protect\citeauthoryear{{Oosterhoff}}{{Oosterhoff}}{1939}]{Oosterhoff}
{Oosterhoff}, P.~T. 1939, The Observatory, 62, 104

\bibitem[\protect\citeauthoryear{{Press} et~al.}{{Press} et~al.}{1992}]{NR}
{Press}, W.~H., {Teukolsky}, S.~A., {Vetterling}, W.~T.,  \& {Flannery}, B.~P.
  1992, {Numerical recipes in C. The art of scientific computing} (Cambridge:
  University Press, |c1992, 2nd ed.)

\bibitem[\protect\citeauthoryear{{Preston}, {Shectman}, \& {Beers}}{{Preston}
  et~al.}{1991}]{Preston91}
{Preston}, G.~W., {Shectman}, S.~A.,  \& {Beers}, T.~C. 1991, \apj, 375, 121

\bibitem[\protect\citeauthoryear{{Reid} \& {Hawley}}{{Reid} \&
  {Hawley}}{2000}]{NLDS}
{Reid}, N.,  \& {Hawley}, S.~L. 2000, {New light on dark stars : red dwarfs,
  low mass stars, brown dwarfs} (Springer-Praxis series in astronomy and
  astrophysics 2000)

\bibitem[\protect\citeauthoryear{{Reimann}}{{Reimann}}{1994}]{Reimann1994}
{Reimann}, J.~D. 1994, Ph.D.~Thesis (University of California)

\bibitem[\protect\citeauthoryear{{Rest}}{{Rest}}{2002}]{Armin_thesis}
{Rest}, A. 2002, Ph.D.~Thesis (University of Washington)

\bibitem[\protect\citeauthoryear{{Robertson} et~al.}{{Robertson}
  et~al.}{2005}]{Robertson2005}
{Robertson}, B., {Bullock}, J.~S., {Font}, A.~S., {Johnston}, K.~V.,  \&
  {Hernquist}, L. 2005, \apj, 632, 872

\bibitem[\protect\citeauthoryear{{Saha}}{{Saha}}{1985}]{Saha85}
{Saha}, A. 1985, \apj, 289, 310

\bibitem[\protect\citeauthoryear{{Sandage}}{{Sandage}}{1993}]{Sandage93}
{Sandage}, A. 1993, \aj, 106, 687

\bibitem[\protect\citeauthoryear{{Sandage}, {Katem}, \& {Sandage}}{{Sandage}
  et~al.}{1981}]{Sandage1981}
{Sandage}, A., {Katem}, B.,  \& {Sandage}, M. 1981, \apjs, 46, 41

\bibitem[\protect\citeauthoryear{{Scargle}}{{Scargle}}{1982}]{Scargle1982}
{Scargle}, J.~D. 1982, \apj, 263, 835

\bibitem[\protect\citeauthoryear{{Schechter}, {Mateo}, \& {Saha}}{{Schechter}
  et~al.}{1993}]{Schechter93}
{Schechter}, P.~L., {Mateo}, M.,  \& {Saha}, A. 1993, \pasp, 105, 1342

\bibitem[\protect\citeauthoryear{{Schlegel}, {Finkbeiner}, \&
  {Davis}}{{Schlegel} et~al.}{1998}]{Schlegel}
{Schlegel}, D.~J., {Finkbeiner}, D.~P.,  \& {Davis}, M. 1998, \apj, 500, 525

\bibitem[\protect\citeauthoryear{{Searle} \& {Zinn}}{{Searle} \&
  {Zinn}}{1978}]{SZ}
{Searle}, L.,  \& {Zinn}, R. 1978, \apj, 225, 357

\bibitem[\protect\citeauthoryear{{Siegel}}{{Siegel}}{2006}]{Siegel2006}
{Siegel}, M.~H. 2006, astro-ph/0607091

\bibitem[\protect\citeauthoryear{{Smith}}{{Smith}}{1995}]{Smith_book}
{Smith}, H.~A. 1995, {RR Lyrae stars} (Cambridge Astrophysics Series,
  Cambridge, New York: Cambridge University Press, 1995)

\bibitem[\protect\citeauthoryear{{Space Telescope Science Institute} \&
  {Osservatorio Astronomico di Torino}}{{Space Telescope Science Institute} \&
  {Osservatorio Astronomico di Torino}}{2001}]{GSC2}
{Space Telescope Science Institute}, .,  \& {Osservatorio Astronomico di
  Torino}. 2001, VizieR Online Data Catalog, 1271, 0

\bibitem[\protect\citeauthoryear{{Stellingwerf}}{{Stellingwerf}}{1978}]{Stelli%
ngwerf1978}
{Stellingwerf}, R.~F. 1978, \apj, 224, 953

\bibitem[\protect\citeauthoryear{{Suntzeff}, {Kinman}, \& {Kraft}}{{Suntzeff}
  et~al.}{1991}]{Suntzeff91}
{Suntzeff}, N.~B., {Kinman}, T.~D.,  \& {Kraft}, R.~P. 1991, \apj, 367, 528

\bibitem[\protect\citeauthoryear{{Venn} et~al.}{{Venn} et~al.}{2004}]{Venn2004}
{Venn}, K.~A., {Irwin}, M., {Shetrone}, M.~D., {Tout}, C.~A., {Hill}, V.,  \&
  {Tolstoy}, E. 2004, \aj, 128, 1177

\bibitem[\protect\citeauthoryear{{Vivas} \& {Zinn}}{{Vivas} \&
  {Zinn}}{2006}]{Vivas2006}
{Vivas}, A.~K.,  \& {Zinn}, R. 2006, astro-ph/0604359

\bibitem[\protect\citeauthoryear{{Vivas}, {Zinn}, et~al.}{{Vivas}
  et~al.}{2004}]{QUEST04}
{Vivas}, A.~K., {Zinn}, R.,  et~al. 2004, \aj, 127, 1158

\bibitem[\protect\citeauthoryear{{Wetterer} \& {McGraw}}{{Wetterer} \&
  {McGraw}}{1996}]{Wetterer96}
{Wetterer}, C.~J.,  \& {McGraw}, J.~T. 1996, \aj, 112, 1046

\bibitem[\protect\citeauthoryear{{Zinn}}{{Zinn}}{1985}]{Zinn1985}
{Zinn}, R. 1985, \apj, 293, 424

\end{thebibliography}





\end{document}